\documentclass[twocolumn,showpacs,aps,prd]{revtex4-1}

\usepackage{graphicx}
\usepackage{dcolumn}
\usepackage{amsmath}
\usepackage{epsfig}
\usepackage{subfigure}
\usepackage{rotating}

\input babarsym

\newcommand{\BABARPubYear}    {11}
\newcommand{\BABARPubNumber}  {027}

\newcommand{\SLACPubNumber} {15135}

\newcommand{\btosgs}{$b\rightarrow s\gamma$\space}
\newcommand{\btosg}{$b\rightarrow s\gamma$}

\def\figurebox#1#2#3{%
    \def\arg{#3}%
    \ifx\arg\empty
    {\hfill\vbox{\hsize#2\hrule\hbox to #2{\vrule\hfill\vbox to #1{\hsize#2\vfill}\vrule}\hrule}\hfill}%
    \else
    {\hfill\epsfbox{#3}\hfill}%
    \fi}

\begin{document}

\preprint{\babar-PUB-\BABARPubYear/\BABARPubNumber}
\preprint{SLAC-PUB-\SLACPubNumber}

\begin{flushleft}
\begin{minipage}{0.45\textwidth}
\babar-PUB-\BABARPubYear/\BABARPubNumber\\
SLAC-PUB-\SLACPubNumber\\
\end{minipage}
\end{flushleft}

\title{
\large \bf
Exclusive Measurements of \btosgs Transition Rate and Photon Energy Spectrum
}

%
\author{J.~P.~Lees}
\author{V.~Poireau}
\author{V.~Tisserand}
\affiliation{Laboratoire d'Annecy-le-Vieux de Physique des Particules (LAPP), Universit\'e de Savoie, CNRS/IN2P3,  F-74941 Annecy-Le-Vieux, France}
\author{J.~Garra~Tico}
\author{E.~Grauges}
\affiliation{Universitat de Barcelona, Facultat de Fisica, Departament ECM, E-08028 Barcelona, Spain }
\author{A.~Palano$^{ab}$ }
\affiliation{INFN Sezione di Bari$^{a}$; Dipartimento di Fisica, Universit\`a di Bari$^{b}$, I-70126 Bari, Italy }
\author{G.~Eigen}
\author{B.~Stugu}
\affiliation{University of Bergen, Institute of Physics, N-5007 Bergen, Norway }
\author{D.~N.~Brown}
\author{L.~T.~Kerth}
\author{Yu.~G.~Kolomensky}
\author{G.~Lynch}
\affiliation{Lawrence Berkeley National Laboratory and University of California, Berkeley, California 94720, USA }
% ...
\collaboration{The \babar\ Collaboration}
\noaffiliation


\begin{abstract}
We use 429\invfb of \epem collision data collected at the \Y4S resonance with 
the \babar\ detector to measure the radiative transition rate of \btosgs with a 
sum of 38 exclusive final states.  The inclusive branching fraction with a 
minimum photon energy of 1.9\gev is found to be 
$\mathcal{B}(\Bbar\rightarrow X_{s}\gamma)=(3.29\pm0.19\pm0.48)\times 10^{-4}$ 
where the first uncertainty is statistical and the second is systematic. We 
also measure the first and second moments of the photon energy spectrum and 
extract the best fit values for the heavy-quark parameters, 
$m_{b}$ and $\mu_{\pi}^{2}$, in the kinetic and shape function models.
\end{abstract}


\maketitle
\section{Introduction}
Flavor changing neutral current processes such as \btosg, forbidden at the tree level in the standard model (SM), occur at leading order through radiative loop diagrams.  Since these diagrams are the dominant contributions to this decay, the effects of many new physics (NP) scenarios, either enhancing or suppressing this transition rate by introducing new mediators within the loop, can be constrained by precision measurements of the total \btosgs transition rate~\cite{2HDM,2HDM2,2HDM3,MFVMSSM,UED}.

In the context of the SM, the first order radiative penguin diagram for the \btosgs transition has a $W$ boson and \t, \c, or \u\ quark in the loop.  The SM calculation for the corresponding \B meson branching fraction has been performed at next-to-next-to-leading order in the perturbative term, yielding $\mathcal{B}(\Bbar\rightarrow X_{s}\gamma)=(3.15\pm0.23)\times 10^{-4}$ for a photon energy of $E_{\gamma}>1.6\gev$, measured in the \B meson rest frame ~\cite{NNLO,NNLO_second}.  Experiments perform this measurement at higher minimum photon energies, generally between 1.7 and 2.0\gev, to limit the background from other \B\ sources. The results are then extrapolated to the lower energy cutoff,  $E_{\gamma}>1.6\gev$, based on different photon spectrum shape functions. The current world average is in good agreement with the SM calculation, and is measured to be $\mathcal{B}(\Bbar\rightarrow X_{s}\gamma)=(3.55\pm0.24\pm0.09)\times 10^{-4},$ for $E_{\gamma}>1.6\gev$~\cite{HFAG}.  The second uncertainty is due to the photon spectrum shape function used to extrapolate to the 1.6 \gev\ photon energy cutoff.

The photon energy spectrum is also of interest, as it gives insight into the momentum distribution function of the \b quark inside the \B meson. Precise knowledge of the function is useful in determining $\left|V_{ub}\right|$ from inclusive semileptonic $\B\rightarrow X_{u}l\nu$ measurements~\cite{Neubert,Wise,Bauer,Ligeti,LNP}.  We fit the measured spectrum to two classes of models: the ``shape function'' scheme~\cite{LNP} and the ``kinetic'' scheme~\cite{BBU}. The photon energy spectra predicted by these models are parameterized to find the best values for the heavy quark effective theory (HQET) parameters, $m_{b}$ and $\mu_{\pi}^{2}$~\cite{Wise}.


Our measurement uses a ``sum of exclusives'' approach, in which we reconstruct the final state of the \s\ quark hadronic system, $X_{s}$, in 38 different modes.  For this article we update a former \babar\ analysis~\cite{babar_measurement} with about five times the integrated luminosity of the previous measurement, as well as an improved analysis procedure.  By reconstructing the $X_{s}$ system, we access the photon energy through:
\begin{equation}
E_{\gamma}^{B} = \frac{m_{B}^{2}-m_{X_{s}}^{2}}{2m_{B}},
\end{equation}

\noindent where $E_{\gamma}^{B}$ is the energy of the transition photon in the \B rest frame, $m_{B}$ is the mass of the \B\ meson, and $m_{X_{s}}$ is the invariant mass of the $X_{s}$ hadronic system.  Measuring $m_{X_{s}}$, with a resolution of around 5\mevcc, gives better resolution on $E_{\gamma}$ than measuring the transition photon directly. We are also able to measure the energy of the transition photon in the rest frame of the \B\ meson rather than correcting for the boost of the \B\ meson with respect to the center of mass (CM) as is required for a direct measurement of the transition photon. We perform this measurement over the range $0.6<m_{X_{s}}<2.8\gevcc$ in 14 bins with a width of 100\mevcc\ for $m_{X_{s}}<2.0\gevcc$, and 4 bins with a width of 200\mevcc\ for $m_{X_{s}}>2.0\gevcc$. To evaluate a total branching fraction for $\mathcal{B}(\Bbar\rightarrow X_{s}\gamma)$ with $E_{\gamma}>1.9\gev$, we sum the partial branching fractions from each $m_{X_{s}}$ bin.  This minimizes our dependence on the underlying photon spectrum structure, and is a departure from our previous procedure~\cite{babar_measurement}, which combined the entire range $0.6 < m_{X_{s}} < 2.8\gevcc$ and used a single fit to the signal yield to determine the total branching fraction.

\section{Detector and Data}

 Our results are based on the entire \Y4S dataset collected with the \babar\ 
detector~\cite{BABARNIM} at the \pep2 asymmetric-energy \B\ factory at the SLAC 
National Accelerator Laboratory.  The data sample has an integrated luminosity 
of 429~\invfb collected at the \Y4S resonance, with a CM energy 
$\sqrt{s}=10.58\gev$, and contains $471\times10^{6}$ \BB\ pairs.  
We refer to this sample as the ``on-peak'' sample.  An ``off-peak'' sample with 
an integrated luminosity of 44.8~\invfb was recorded about 40~\mev below 
the \Y4S resonance, and is used for the study of backgrounds consisting of 
\epem production of light \qqbar (\q=\u, \d, \s, \c) or \tautau states.

The \babar\ detector is described in detail in~\cite{BABARNIM}.  Charged-particle momenta are measured by the combination of a silicon vertex tracker (SVT), consisting of five layers of double-sided silicon strip detectors, and a 40-layer central drift chamber (DCH) having a combination of axial and stereo wires.

Charged-particle identification is provided by the combination of the average energy loss ($dE/dx$) measured in the tracking devices and the Cherenkov-radiation information measured by an internally reflecting ring-imaging Cherenkov detector (DIRC).

Photon and electron energies are measured by a CsI(Tl) electromagnetic calorimeter (EMC). The SVT, DCH, DIRC, and EMC operate inside of a 1.5 T magnet.  Charged $\pi$/$\mu$ separation is done using the instrumented flux return of the magnetic field, originally instrumented with resistive plate chambers~\cite{BABARNIM} and later with limited streamer tubes~\cite{LST}.

\section{Signal and Background Simulation}\label{sec:simulation}

To avoid experimental biases, we use Monte Carlo (MC) simulations to model both the expected signal and background events and to define selection criteria before looking at the data.  We have produced MC samples for $\epem\rightarrow\qqbar$ (\q=\u,\d,\s,\c) and $\epem\rightarrow\tautau$ events, each at two times the on-peak luminosity, as well as \BB\ MC events, excluding decays of the \B meson to an $X_{s}\gamma$ final state, at three times the on-peak luminosity. We also consider ``cross-feed'' backgrounds.  We define cross-feed as signal events in which we wrongly reconstruct the \B\ candidate. This occurs either because the $X_{s}$ final state is not one of the 38 reconstructed modes, not all of the particles in the true final state are detected, or the procedure for selecting the correctly reconstructed \B\ from several potential \B\ candidates fails in some cases.

Two types of signal MC events are generated, one for the $K^{*}(892)$ region ($m_{X_{s}}<1.1\gevcc$) in which the \btosgs transition proceeds exclusively through $B\rightarrow K^{*}(892)\gamma$, and one for the region above the $K^{*}(892)$ resonance ($1.1<m_{X_{s}}<2.8\gevcc$, the upper bound being the limit of our ability to adequately reject \B\ backgrounds).  While there are several known $X_{s}$ resonances above the $K^{*}(892)$, and evidence for even more~\cite{PDG}, these resonances are broad and overlapping.  We therefore take only the $K^{*}(892)$ resonance explicitly into account when simulating the signal events, as recommended by~\cite{KN}.

The quarks in inclusive region signal MC events shower using the ``phase-space hadronization model,'' as opposed to the well known ``Lund string model,'' with our default JETSET~\cite{JETSET} settings.  The most important JETSET parameters that influence the fragmentation of the $X_{s}$ system in this inclusive region are the probabilities of forming a spin-1 state for the \s\ quark or \u/\d\ quarks (the corresponding JETSET parameters are \texttt{parj(12)} and \texttt{parj(11)}).  These probabilities are set to 0.60 and 0.40, respectively.

We generate the inclusive signal MC events with a flat photon spectrum with bounds corresponding to the $m_{X_{s}}$ boundaries, which we then reweight to match whichever spectrum model we choose.  We do not take any explicit photon model into account when evaluating signal efficiency within a given $X_{s}$ mass bin.  However, to evaluate the optimal background-rejection requirements, we do need to specify the expected shape of the spectrum.  For this, we use the model settings for the kinetic scheme models~\cite{BBU} found to be consistent with the previous \babar\ sum of exclusive analysis ($m_{b}=4.65\gevcc,\ \mu_{\pi}^{2}=0.20\gev^{2}$)~\cite{babar_measurement}.

GEANT4~\cite{geant} is used to model the response of the detector for all MC samples. Time-dependent detector inefficiencies, monitored during data taking, are also included. 
\section{\bf{$B$} meson Reconstruction and Background Rejection}

We reconstruct the \B meson in one of 38 final states of the $X_{s}$ plus a high energy photon,
as listed in Table~\ref{tab:modes}~\cite{conjugates}.  These modes consist of one or three kaons, at most
one $\eta$, and at most four pions, of which no more than two can be neutral pions.  The method of particle identification
(PID) has improved over the run of the experiment. In particular for charged $K$
identification, we use a multi-class classifier procedure of error correcting output code (ECOC)~\cite{ECOC}.  The kaon identification
efficiency is roughly 90\% for the momentum range considered for this analysis.

\begin{table}[htp]
\begin{small}
\begin{center}
\caption{\label{tab:modes} The 38 $X_{s}$ decay modes used for \B meson reconstruction in this analysis.}
\begin{tabular}{clcl}
\hline
\hline
Mode Num.		& Final State							&Mode Num.	&Final State\\
\hline
1		& $ \KS\pi^{+}$				& 20		& $ \KS\pi^{+}\pi^{-}\pi^{+}\pi^{-}$\\
2		& $ K^{+}\pi^{0}$				& 21 		& $ K^{+}\pi^{+}\pi^{-}\pi^{-}\pi^{0}$\\
3		& $ K^{+}\pi^{-}$				& 22 		& $ \KS\pi^{+}\pi^{-}\pi^{0}\pi^{0}$\\
4		& $ \KS\pi^{0}$				& 23		& $ K^{+}\eta$\\
5		& $ K^{+}\pi^{+}\pi^{-}$			& 24 		& $ \KS\eta$\\
6		& $ \KS\pi^{+}\pi^{0}$			& 25 		& $ \KS\eta\pi^{+}$\\
7		& $ K^{+}\pi^{0}\pi^{0}$			& 26		& $ K^{+}\eta\pi^{0}$\\
8		& $ \KS\pi^{+}\pi^{-}$			& 27		& $ K^{+}\eta\pi^{-}$\\
9		& $ K^{+}\pi^{-}\pi^{0}$			& 28		& $ \KS\eta\pi^{0}$\\
10		& $ \KS\pi^{0}\pi^{0}$			& 29		& $ K^{+}\eta\pi^{+}\pi^{-}$\\
11		& $ \KS\pi^{+}\pi^{-}\pi^{+}$		& 30 		& $ \KS\eta\pi^{+}\pi^{0}$\\
12		& $ K^{+}\pi^{+}\pi^{-}\pi^{0}$		& 31		& $ \KS\eta\pi^{+}\pi^{-}$\\
13		& $ \KS\pi^{+}\pi^{0}\pi^{0}$		& 32		& $ K^{+}\eta\pi^{-}\pi^{0}$\\
14		& $ K^{+}\pi^{+}\pi^{-}\pi^{-}$		& 33		& $ K^{+}K^{-}K^{+}$\\
15		& $ \KS\pi^{0}\pi^{+}\pi^{-}$		& 34		& $ K^{+}K^{-}\KS$\\
16		& $ K^{+}\pi^{-}\pi^{0}\pi^{0}$		& 35		& $ K^{+}K^{-}\KS\pi^{+}$\\
17		& $ K^{+}\pi^{+}\pi^{-}\pi^{+}\pi^{-}$	& 36		& $ K^{+}K^{-}K^{+}\pi^{0}$\\
18		& $ \KS\pi^{+}\pi^{-}\pi^{+}\pi^{0}$	& 37		& $ K^{+}K^{-}K^{+}\pi^{-}$\\
19		& $ K^{+}\pi^{+}\pi^{-}\pi^{0}\pi^{0}$	& 38		& $ K^{+}K^{-}\KS\pi^{0}$\\
\hline
\hline
\end{tabular}
\end{center}
\end{small}
\end{table}

The \KS mesons are reconstructed as $\KS\to\pi^{+}\pi^{-}$ candidates with an invariant $\pi^{+}\pi^{-}$ mass within 9\mevcc of the
nominal \KS mass~\cite{PDG}, a flight distance greater than 0.2 cm from the primary event vertex,
and a flight significance (measurement of flight distance divided by the uncertainty on the measured distance)
greater than 3.  We do not include \KL mesons or $\KS\to\pi^{0}\pi^{0}$ decays in our reconstructed final states.

Charged $K$ candidates are identified based on the ECOC algorithms~\cite{ECOC}, which use
information from the tracking system, the DIRC, and the EMC to identify particle
species using multivariate classifiers.  All remaining charged tracks are assumed to originate from charged pions.

The $\pi^{0}$ and $\eta$ candidates are reconstructed from photon candidates with an energy greater
than $60\mev$ as measured in the laboratory frame, and must have an invariant mass between $115$
and $150\mevcc$ for the $\pi^{0}$, and $470$ and $620\mevcc$ for the $\eta$.  We also require
a minimum momentum $p_{\pi^{0},\eta}>200\mevc$ in the lab frame.  Although we do not explicitly
reconstruct the $\eta\rightarrow\pi^{+}\pi^{-}\pi^{0}$ decay mode, this mode is implicitly included in the final
states if there is at most one other pion in the event.  We combine these charged and neutral particles
to form different $X_{s}$ candidates in the event.

We require that an event contain at least one photon candidate with $1.6<E^{*}_{\gamma}<3.0\gev$
(where ``*'' henceforth indicates variables measured in the CM), which is consistent
with the signal photon of the decay \btosg.  The distance to the closest cluster in
the EMC is required to be greater than 25 cm from this signal photon cluster.  We also require the
angle between the signal photon candidate and the thrust axis 
of the rest of the event to satisfy
$\left|\cos{\theta_{T\gamma}^{*}}\right|<0.85$, and the ratio of event shape angular
moments to satisfy $L_{12}/L_{10}<0.46$~\cite{legendre} (the signal peaks at slightly lower values than the background).
These two preliminary requirements on the event topology are
especially effective at reducing the large amount of more jet-like light $q \bar{q}$\ backgrounds, and
together decrease this background source by about $50\%$ (while only removing $10\%$ of signal).

We combine the $X_{s}$ candidates and the signal photon candidates to form \B candidates in the event.
We define the beam-energy substituted mass, $\mes=\sqrt{(\sqrt{s}/2)^{2}-(p_{B}^{*})^{2}}$, and
require $\mes>5.24\gevcc$.  We also require the difference between the expected \B energy and the reconstructed \B energy,
$\left|\Delta E\right|=\left|E_{B}^{*}-\sqrt{s}/2\right|$, to satisfy $\left|\Delta E\right|<0.15\gev$.  For these quantities,
$p_{B}^{*}$ and $E_{B}^{*}$ are the momentum and energy of the reconstructed \B meson in the CM system.

With these loose preliminary requirements in place, each event still typically has several \B meson candidates.
We construct a random forest classifier~\cite{RF} (a signal selecting classifier, SSC) to find the best candidate
in an event.  This classifier is built using the variables $\Delta E/\sigma_{E}$ (where $\sigma_{E}$ is the uncertainty
on the total energy of the reconstructed \B), the thrust of the reconstructed \B, the $\pi^{0}$ momentum in the CM
(if the candidate has a $\pi^{0}$), the invariant mass of the $X_{s}$ candidate, and the zeroth and fifth Fox-Wolfram moments of
the event~\cite{FW}.  We choose to include the fifth Fox-Wolfram moment because our MC simulation indicates that this variable
improves the performance of our classifier.
The selected \B candidate in an event is the candidate with the highest response to this classifier.
We find that applying this classifier to select the best candidate, after placing a loose requirement
on $\left|\Delta E\right|$, rather than selecting the candidate with the smallest $\left|\Delta E\right|$, improves
the signal efficiency by
a factor of about two. We also find that placing a requirement on the SSC response is effective at further removing
\B backgrounds.

To further reduce the background from events in which a photon from a high energy $\pi^{0}$ decay is mistaken as the
signal photon candidate, we construct a dedicated $\pi^{0}$ veto using a random forest classifier~\cite{RF}.  If the signal
photon candidate in an event can be combined with any other photon to form a candidate with an invariant mass in
the range $115<m_{\gamma\gamma}<150\mevcc$, we evaluate the $\pi^{0}$ veto classifier response based on the invariant
mass of the two photons and the energy of the lower energy photon. 
The response of the $\pi^{0}$ veto classifier is used as input to a more general background rejecting classifier (BRC).

The BRC is constructed to remove continuum (lighter $q \bar{q}$) backgrounds. To construct this classifier, we use
information from the $\pi^{0}$ veto, $\left|\cos{\theta_{T\gamma}^{*}}\right|$, $\left|\cos{\theta_{T}^{*}}\right|$
(the angle between the thrust axis of the \B and the thrust axis of the rest of the event),
$\left|\cos{\theta_{B}^{*}}\right|$ (the CM polar angle of the \B flight direction), the zeroth, first, and
second angular moments~\cite{legendre} computed along the signal photon candidate's axis as well as the ratio $L_{12}/L_{10}$ (which exhibits
slightly different signal and background shapes), and
the $10^{\circ}$ momentum flow-cones around the \B flight-direction.

To effectively remove background while maintaining signal efficiency, we evaluate
optimal requirements for the responses of the BRC and SSC in four mass
regions, [0.6--1.1], [1.1--2.0], [2.0--2.4], and [2.4--2.8]\gevcc, optimizing
the figure of merit $S/\sqrt{S+B}$, where $S$ is the expected signal yield and $B$ is the expected background yield evaluated using MC simulation.

\section{Signal Yield Extraction}\label{sec:fitting}

We extract the signal yield by performing fits to the \mes distribution in each bin of $m_{X_{s}}$.  The signal distribution is described by a Crystal Ball function (CB)~\cite{CB}:
\begin{small}
\begin{align}
\nonumber f(\mes) = e^{\left(-\frac{(\mes-m_{0})^{2}}{2\sigma^{2}}\right)}, \quad 
\left|\frac{\mes-m_{0}}{\sigma}\right| < \alpha,\\
f(\mes) = \frac{\left(\frac{n_{\rm CB}}{\alpha}\right)^{n_{\rm CB}}e^{\left(-\frac{\alpha^{2}}{2}\right)}}{\left(\frac{n_{\rm CB}}{\alpha}-\alpha-\frac{\mes-m_{0}}{\sigma}\right)^{n_{\rm CB}}}, \quad
\left|\frac{\mes-m_{0}}{\sigma}\right| > \alpha,
\end{align}
\end{small}
\noindent
where $m_{0}$ and $\sigma$ are the peak  position and width, respectively, and the parameters $\alpha$ and $n_{\rm CB}$ take account of the non-Gaussian tail.  This distribution takes into account the asymmetry of the $\mes$ distribution for these events. The backgrounds are described by ARGUS functions~\cite{argus} for the combinatorial components:
\begin{equation}
f(\mes) = \mes\left(1-\left(\frac{\mes}{m}\right)^{2}\right)^{\frac{1}{2}}\times e^{\left(c\frac{\mes}{m}\right)},
\end{equation}
\noindent where $m$ is the end point, and $c$ is the slope, and Novosibirsk functions~\cite{Nvs} for both the peaking \BB\ contribution and peaking cross-feed contribution (``peaking'' meaning apparently resonant behavior similar to the signal distribution in $\mes$).

The signal CB distribution is parameterized based on a fit to correctly reconstructed signal MC events over the full hadronic mass range, 0.6--2.8\gevcc, as we find little $X_{s}$ mass dependence of the signal shape parameters.  The CB parameters take the values $\alpha$=1.12, $m_{0}$=5.28 $\gevcc$, $\sigma$=2.84 $\mevcc$, and $n_{\rm CB}=145$ for every mass bin.  
In Sec.~\ref{sec:systematics} we evaluate the uncertainties indroduced by fixing the CB shape parameters.

The cross-feed shape has both a peaking component and a combinatoric tail.  The peaking component is described by a Novosibirsk function, parameterized over five different mass regions, [0.6--1.1], [1.1--1.5], [1.5--2.0], [2.0--2.4], and [2.4--2.8]\gevcc, based on MC distributions over these regions.  
The combinatoric cross-feed tail is described by an ARGUS function with the slope $c$ fit to the MC events in each mass bin, and fixed in the fits to data.  We fix the fraction of peaking cross-feed MC events, the fraction of signal to signal+cross-feed events, and the shapes of the cross-feed Novosibirsk and ARGUS functions, in each bin of $m_{X_{s}}$, based on the MC events. We allow the total signal+cross-feed yield to float in each mass bin in the fits to data.

A second ARGUS function is used to parameterize the combinatoric background from 
continuum and other \BB\ sources.  We fix the end point $m$ of the ARGUS function to the kinematic limit (5.29\gevcc) of the $\mes$ variable and allow the yield to float.

The \BB\ background also has a peaking component, which becomes more significant at higher $X_{s}$ mass, is also described by a Novosibirsk function, and is parameterized over three mass ranges, [0.6--2.0], [2.0--2.4], and [2.4--2.8]\gevcc.  We fix the total number and shape of the peaking \BB\ events based on a fit to the \BB\ MC events in each mass bin.

\begin{figure}[htp]
\begin{center}
\includegraphics[width=0.4\textwidth]{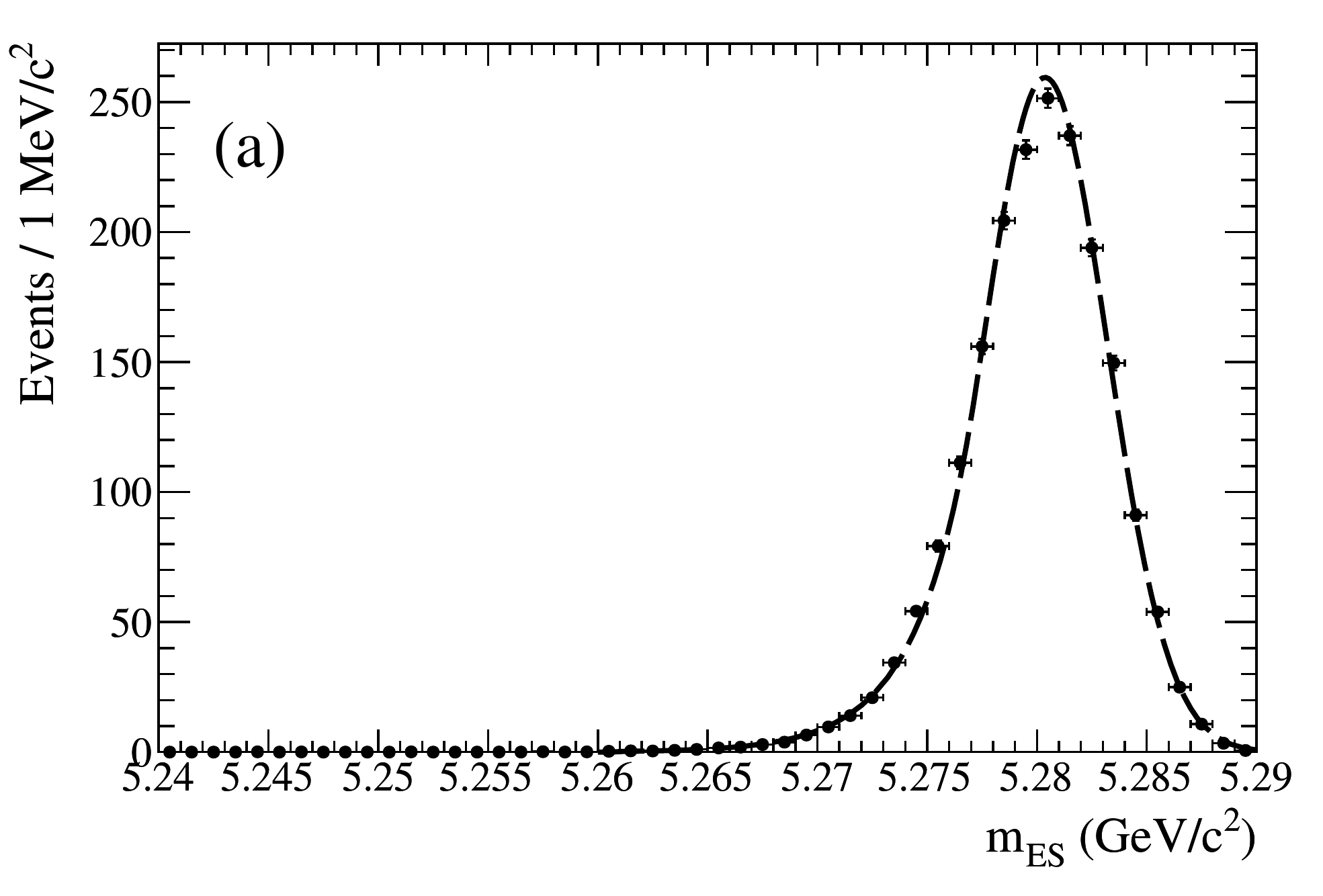}\\
\includegraphics[width=0.4\textwidth]{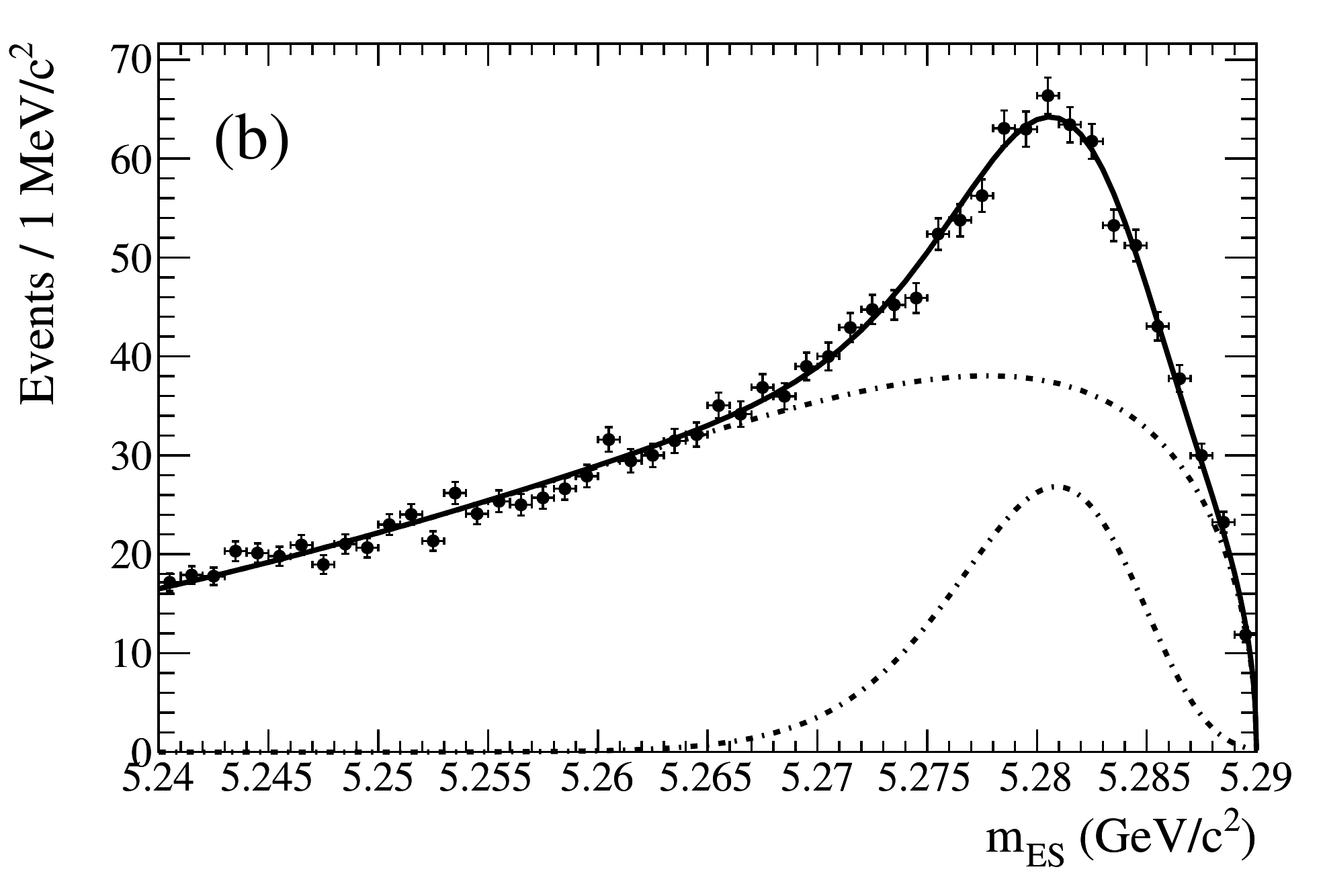}\\
\includegraphics[width=0.4\textwidth]{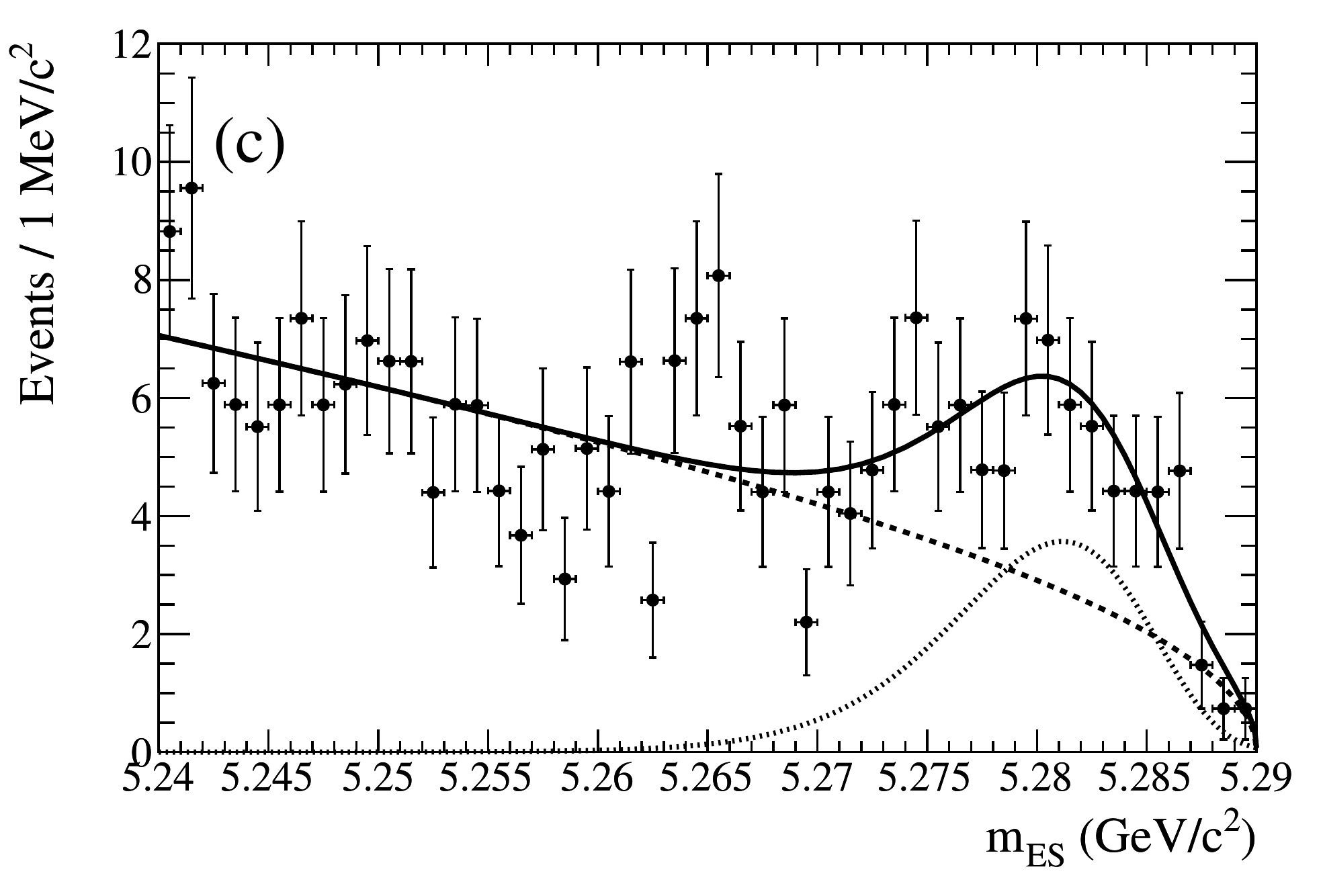}\\
\includegraphics[width=0.4\textwidth]{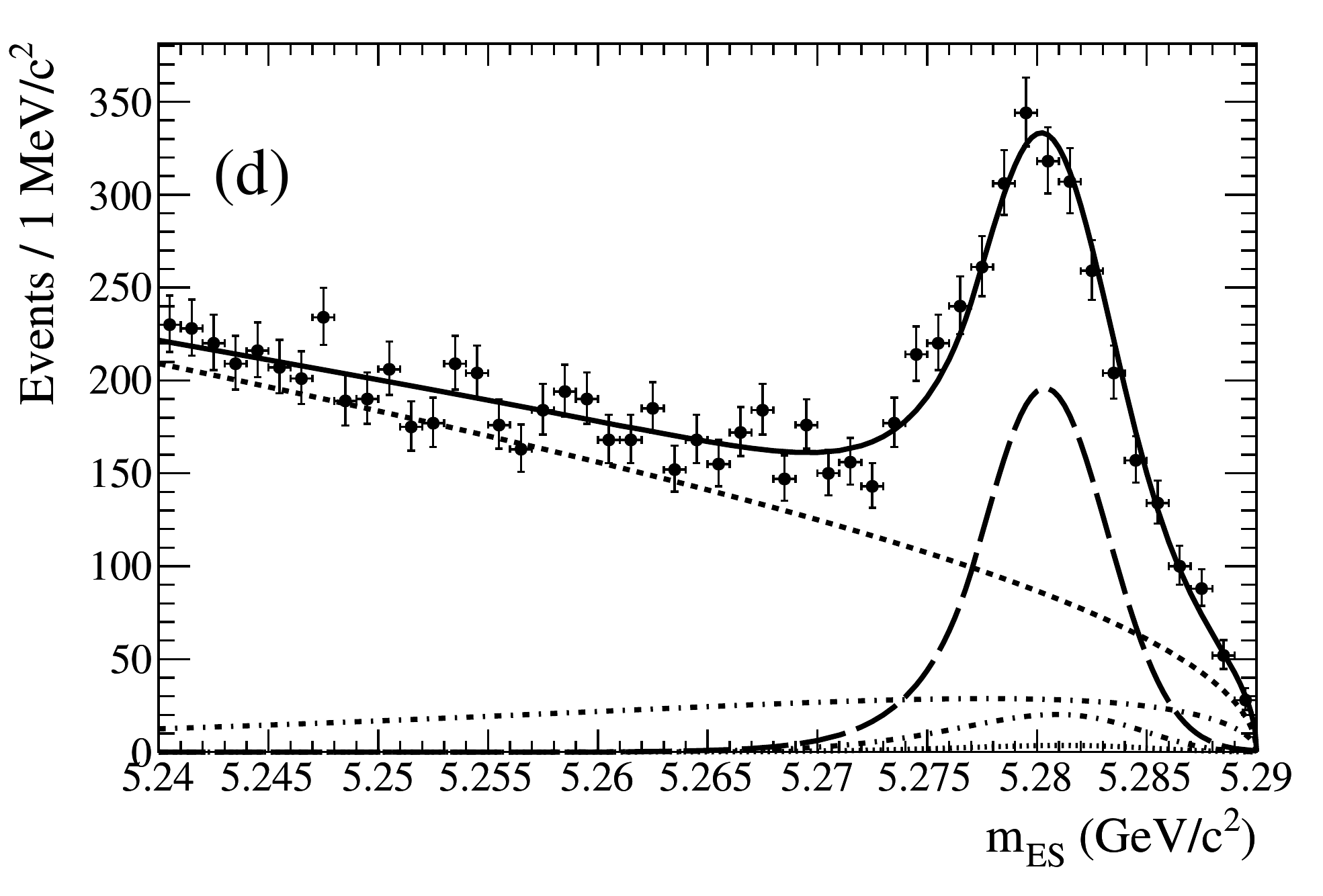}
\caption{The fit in mass bin $1.4<m_{X_{s}}<1.5$\gevcc\ to (a) signal MC events, (b) cross-feed MC events, (c) peaking \BB MC events, and (d) the data. The signal (thick dashed), cross-feed (two dot-dashed curves, one ARGUS function and one Novosibirsk function), peaking \BB (dotted), and combinatoric background (thin dashed) component functions are shown.\label{fig:fit_plot}}
\end{center}
\end{figure}

We perform a minimum $\chi^{2}$ fit to the $\mes$ distribution in each bin of $m_{X_{s}}$, allowing the slope of the combinatoric ARGUS and the fractional yield of signal+cross-feed to float (the complementary fractional yield, once the peaking \BB\ is accounted for, reflects the normalization of the combinatoric ARGUS function). Figure~\ref{fig:fit_plot} shows an example for $m_{X_{s}}$ bin 1.4--1.5\gevcc.  We fix all other shape parameters, and evaluate systematic uncertainties associated with fixing these parameters in Sec.~\ref{sec:systematics}.  We perform MC simulations (``toy studies'') to ensure that we do not introduce any biases due to the fitting procedure.

\section{$X_{s}$ Fragmentation and Missing Fraction}\label{sec:fragmentation}

The fragmentation of the hadronic system in the inclusive region, $1.1<m_{X_{s}}<2.8\gevcc$, is modeled with JETSET with a phase-space hadronization model.  The differences between fragmentation in the MC sample and in the data influence the measurement in two ways. First, since the efficiencies for the 38 modes are not the same, an incorrect modeling of their relative fractions will lead to an incorrect expected total efficiency for reconstructing the 38 final states ($\epsilon_{38}$).  Second, the simulation of the fragmentation process can introduce incorrect estimates of the fraction of the total inclusive \btosgs transition rate reflected by the 38 modes ($\epsilon_{\mathrm{incl}}$).  The fraction of final states in each of the mass bins that is not included in our 38 modes is referred to as the ``missing fraction'', and is equivalent to $1-\epsilon_{\mathrm{incl}}$.

We are able to evaluate and correct $\epsilon_{38}$ for the first effect, and use these results to estimate the uncertainty on the second effect, our uncertainty on $\epsilon_{\mathrm{incl}}$, by performing a fragmentation study comparing the frequency of groups of modes in the MC sample to the data.  For this study, we compare the frequency of ten groups of modes, each containing two to ten final states, in the MC sample to the frequency for these groups found in the data.  We perform this study in four different mass regions, [1.1--1.5], [1.5--2.0], [2.0--2.4], and [2.4--2.8]\gevcc.

The procedure for the study involves reweighting the relative contribution of each of the groups of modes in our MC based on the relative amount found in the data. The efficacy of the procedure is checked on MC events by ensuring we can find the $\epsilon_{38}$ in each mass bin for the Lund string model when starting with the default phase-space hadronization model~\cite{JETSET}, as well as find the $\epsilon_{38}$ in each mass bin for the phase-space hadronization model when starting with the Lund string model. The different groups of modes we use to compare data and the MC samples, along with the results of the comparisons in each mass bin, are given in Table~\ref{tab:ratios}, obtained with the default phase-space hadronization model as the starting point.
\begin{table*}[htp]
\begin{center}
\caption{\label{tab:ratios} The subsets of modes and the ratio of yields found in each $m_{X_{s}}$ region when comparing the data to the MC events.  The error is statistical only.}
\begin{tabular}{ccccc}
\hline
\hline
Data 		& \multicolumn{3}{c}{Definition}			& Modes used			\\
subset		& \multicolumn{3}{c}{}				&				\\
\hline
1			& \multicolumn{3}{c}{2 bodies without $\pi^{0}$}	&  1,3			\\
2			& \multicolumn{3}{c}{2 bodies with 1 $\pi^{0}$}	&  2,4			 \\
3			& \multicolumn{3}{c}{3 bodies without $\pi^{0}$}	&  5,8		\\
4			& \multicolumn{3}{c}{3 bodies with 1 $\pi^{0}$}	&  6,9			\\
5			& \multicolumn{3}{c}{4 bodies without $\pi^{0}$}	&  11,14	\\
6			& \multicolumn{3}{c}{4 bodies with 1 $\pi^{0}$}	&  12,15		\\
7			& \multicolumn{3}{c}{3/4 bodies with 2 $\pi^{0}$}	&  7,10,13,16		\\
8			& \multicolumn{3}{c}{5 bodies with 0-2 $\pi^{0}$}	&  17-22		\\
9			& \multicolumn{3}{c}{$\eta\rightarrow\gamma\gamma$}	&  23-32		\\
10			& \multicolumn{3}{c}{3K modes}			&  33-38		\\

\hline
\hline
Data    &$1.1 < m_{X_s} < 1.5\gevcc$     &$1.5 < m_{X_s} < 2.0\gevcc$     &$2.0 < m_{X_s} < 2.4\gevcc$     &$2.4 < m_{X_s} < 2.8\gevcc$\\
subset  & (data/MC)     &(data/MC)&(data/MC)&(data/MC)\\
\hline
1   & 0.65 $\pm$ 0.03   &0.38 $\pm$ 0.03    &0.05$\pm$0.05      &0.18$\pm$ 0.13\\
2   &0.53 $\pm$ 0.05    &0.28 $\pm$ 0.06    &0.32 $\pm$ 0.12    &0.15 $^{+0.25}_{-0.15}$ \\
3   &1.20 $\pm$ 0.03    &1.01 $\pm$ 0.04    &0.72 $\pm$ 0.11    &0.25$\pm0.25$ \\
4   &1.70 $\pm$ 0.05    &1.03 $\pm$ 0.06    &0.33 $\pm$ 0.13    &1.00$^{+0.47}_{-1.00}$ \\
5   &0.34 $\pm$ 0.08    &1.34 $\pm$ 0.10    &1.12 $\pm$ 0.23    &2.29 $\pm$ 0.74\\
6   &1.24 $\pm$ 0.13    &1.16 $\pm$ 0.11    &1.28 $\pm$ 0.27    & 0.10$^{+0.39}_{-0.10}$\\
7   &0.56 $\pm$ 0.19    &1.37 $\pm$ 0.30    &0.83 $\pm$ 0.53    &2.06 $\pm$ 1.64\\
8   &1.00$^{+1.05}_{-1.00}$&0.57 $\pm$ 0.16 &0.74 $\pm$ 0.28    &0.29$^{+1.27}_{-0.29}$\\
9   &0.94 $\pm$ 0.15    &1.70 $\pm$ 0.20    &2.47 $\pm$ 0.50    &1.09$^{+1.03}_{-1.09}$ \\
10  &0.00 $\pm$ 0.00    &0.62 $\pm$ 0.11    &0.74 $\pm$ 0.31    &0.83$^{+1.11}_{-0.83}$ \\
\hline
\hline
\end{tabular}
\end{center}
\end{table*}

To perform this study, we combine the mass bins into the four mass regions, and fit the signal+crossfeed contribution for each subset of modes in each mass region in the data.  We then use the ratio of the yield of each subset found in data to the amount found in the MC sample to reweight the MC sample to better reflect the data in the mass region.  
We use the statistical uncertainty in fitting each subset in data as the uncertainty on the ratio.

After correcting the signal and cross-feed MC events based on these comparisons, we evaluate the value of $\epsilon_{38}$ for each mass bin, reported in Table~\ref{tab:eps38}. For the inclusive region, the uncertainty on $\epsilon_{38}$ is calculated using the
uncertainties in the fragmentation corrections, as described later in Sec.~\ref{sec:systematics}.  Since the fragmentation in the $K^{*}(892)$ region is considered well modeled, we do not perform a fragmentation correction on these mass bins.

\begin{table}[htp]
\begin{center}
\caption{\label{tab:eps38}The value of $\epsilon_{38}$ before and after the fragmentation corrections are performed on the inclusive region.  The uncertainty on the corrected value in the inclusive region reflects the uncertainty of the fits to the data.}
\begin{tabular}{ccc}
\hline
\hline
$m_{X_{s}}$	& $\epsilon_{38}$ original	& $\epsilon_{38}$ final	\\
	(\gevcc)	& (\%)				& (\%)				\\
\hline
0.6--0.7		& 15.0				&15.0\\
0.7--0.8		& 16.5				&16.5\\
0.8--0.9		& 17.3				&17.3\\
0.9--1.0		& 18.3				&18.3\\
1.0--1.1		& 16.0				&16.0\\
1.1--1.2		& 11.5				&10.4$\pm$0.4\\
1.2--1.3		& 11.6				&10.6$\pm$0.3\\
1.3--1.4		& 10.7				&9.9$\pm$0.3\\
1.4--1.5		& 9.5				&8.9$\pm$0.5\\
1.5--1.6		& 8.4				&7.5$\pm$0.5\\
1.6--1.7		& 7.2				&6.5$\pm$0.4\\
1.7--1.8		& 5.5				&5.0$\pm$0.4\\
1.8--1.9		& 4.5				&4.2$\pm$0.4\\
1.9--2.0		& 3.3				&3.0$\pm$0.4\\
2.0--2.2		& 4.0				&3.2$\pm$0.4\\
2.2--2.4		& 3.1				&2.4$\pm$0.4\\
2.4--2.6		& 2.3				&1.9$\pm$0.7\\	
2.6--2.8		& 2.3				&2.1$\pm$0.9\\
\hline
\hline
\end{tabular}
\end{center}
\end{table}

We base the uncertainty on the fraction of the inclusive \btosgs transitions measured by the 38 final states, $\epsilon_{\mathrm{incl}}$, on the range of values predicted by competing fragmentation models in the MC samples. We consider many settings of JETSET using both the default phase-space and the Lund string hadronization mechanism, as well as a thermodynamical model~\cite{QR}.  
Other models in JETSET (Field-Feynman model~\cite{Fey} of the showering quark system, etc.) are found to yield results consistent with the Lund string model, and are not further considered.

As mentioned above, we identify the probabilities for forming a spin-1 hadron with the \s quark or \u/\d quarks to be the JETSET parameters that have the largest impact on the breakdown of final states.  We try many settings for these parameters in both the phase-space hadronization mechanism and the Lund string model mechanism in JETSET.  By varying the spin-1 probabilities and using both of these fragmentation mechanisms, we are able to identify a range of models that, taken together, account for the breakdown of final states found in the data in the fragmentation study (Table~\ref{tab:ratios}).  We vary the probability for forming a spin-1 hadron with the \s quark between zero and one, and with the \u/\d quark between 0.2 and 0.8. When comparing to our default MC settings, the models we consider predict both higher and lower ratios than those found in the data, but no single model matches every ratio in every mass region.

We also find that no single mechanism or JETSET setting perfectly reproduces the fragmentation in the data; however the models chosen bound the data.  The fact that spin-1 probability settings need to be varied to account for data and MC differences is expected, as a variety of resonances exist in the inclusive region. 
The maximum, minimum, and default values for $\epsilon_{\mathrm{incl}}$ that we find are reported in Table~\ref{tab:incl}.  We account for what is seen in data in the fragmentation study through a variety of settings of both the Lund string mechanism and phase-space hadronization mechanism, and therefore base our uncertainty on $\epsilon_{\mathrm{incl}}$ on these same models.\
The statistics model and the default JETSET settings predict values for $\epsilon_{\mathrm{incl}}$ that lie between those predicted by other settings of JETSET that we try.  As we find that no model exactly describes the fragmentation we observe in the data, but together the models considered bound the data, we count each model as equally probable, and take the systematic uncertainty on the correct value for $\epsilon_{\mathrm{incl}}$ as the difference between the maximum and minimum values of $\epsilon_{\mathrm{incl}}$ relative to the default MC value, and divide by $\sqrt{12}$, reflecting the standard deviation of a uniform distribution.

\begin{table}[htp]
\begin{center}
\caption{The minimum, maximum, and default values of $\epsilon_{\mathrm{incl}}$ found for the range of models that account for the differences seen between the default MC events and data in the inclusive region. We include the $K^{*}(892)$ region default values as well, though these mass bins are not modeled by the inclusive MC sample.\label{tab:incl}}
\begin{tabular}{cccc}
\hline
\hline
$m_{X_{s}}$     &Minimum    & Maximum   &Default \\
(\gevcc)        &$\epsilon_{\mathrm{incl}}$   &$\epsilon_{\mathrm{incl}}$   &$\epsilon_{\mathrm{incl}}$\\
\hline
0.6--0.7         & --        & --        & 0.75\\
0.7--0.8         & --        & --        & 0.74\\
0.8--0.9         & --        & --        & 0.74\\
0.9--1.0         & --        & --        & 0.75\\
1.0--1.1         & --        & --        & 0.74\\
1.1--1.2         & 0.71      & 0.74      & 0.73\\
1.2--1.3         & 0.71      & 0.74      & 0.72\\
1.3--1.4         & 0.70      & 0.74      & 0.72\\
1.4--1.5         & 0.69      & 0.73      & 0.71\\
1.5--1.6         & 0.66      & 0.73      & 0.68\\
1.6--1.7         & 0.59      & 0.72      & 0.66\\
1.7--1.8         & 0.57      & 0.72      & 0.63\\
1.8--1.9         & 0.52      & 0.71      & 0.59\\
1.9--2.0         & 0.47      & 0.68      & 0.54\\
2.0--2.2         & 0.41      & 0.64      & 0.48\\
2.2--2.4         & 0.33      & 0.60      & 0.39\\
2.4--2.6         & 0.27      & 0.56      & 0.31\\
2.6--2.8         & 0.23      & 0.51      & 0.25\\
\hline
\hline
\end{tabular}
\end{center}
\end{table} 
\section{Systematic Uncertainties}\label{sec:systematics}

We present the $X_{s}$ mass-bin-dependent uncertainties in Table~\ref{tab:systematic}.
The uncertainty on the total number of \B mesons produced at \babar\ is evaluated at 1.1\%.

The uncertainty on the efficiency of the requirements on the two multivariate classifiers are evaluated in signal-like data sidebands, regions in parameter space similar to, but not overlapping with, the signal region, by comparing the efficiency of the requirements on MC events and the efficiency of these requirements on data.  We define our sidebands as the inverse of the requirements we place on the classifiers. Therefore if we require the SSC response to be greater than 0.5 for a mass region, we evaluate the BRC uncertainty in the SSC sideband defined by requiring an SSC response less than 0.5 (and similarly for evaluating the SSC uncertainty in the BRC-defined sideband).  The relative difference between the two efficiencies is taken as the systematic uncertainty.  The sideband produced by taking the inverse of the requirements on the SSC is used to evaluate the uncertainty on the requirements on the BRC.

To evaluate the uncertainty on the SSC requirement, the events that are identified by the $\pi^{0}$-veto classifier to contain a $\pi^{0}$ candidate are used with the further requirement $m_{ES}>5.27\gevcc$.  This gives a more signal-like sample of events that have a high energy $\pi^{0}$ in place of the signal transition photon.  The efficiency of the SSC requirement is compared between data and the MC events with the use of this sideband.

To evaluate fitting uncertainties related to fixing many of the parameters in the signal and cross-feed PDFs, we use the $K^{*}(892)$ region ($m_{X_{s}}<1.1\gevcc$) to determine reasonable shifts in these parameters.   We assign the systematic uncertainty as the change in signal yield in the fit to data when we use the shifted shape parameters.  For the uncertainty on the fraction of signal to signal+cross-feed, which is also fixed in the fit to data, we fix the total yield and slope of this ARGUS function (these are the two parameters that we float in the fits to data) and allow this fraction to float in each mass bin.  We take the change in signal yield when we fix the signal fraction to this new value as the systematic uncertainty.

To evaluate the uncertainty on the peaking \BB background PDF shape, we use the change in signal yield when changing the parameter values by the uncertainty in the fits to MC events. 

The uncertainty on the number of peaking \BB\ events, generally the largest source of \BB fitting error in Table~\ref{tab:systematic}, is again evaluated based on the $\pi^{0}$-veto sideband.  In this sideband, we evaluate the \BB\ MC predictions for the number of peaking events and compare this to the number of peaking \BB\ events we find in data.  We find these values to agree within one standard deviation for the three mass regions over which we have parameterized the peaking \BB\ Novosibirsk function (see Sec.~\ref{sec:fitting}).  We determine the mass-region-dependent uncertainty on the measurement of peaking \BB\ yield in the $\pi^{0}$-veto sideband in data.  We use this uncertainty added in quadrature with the uncertainty from the fits to the \BB\ MC sample as the uncertainty on the number of peaking \BB\ events in each mass bin. Unlike the other systematic uncertainties, which are multiplicative in nature, this uncertainty is additive since we are subtracting out peaking \BB\ events we would otherwise fit as signal+cross-feed in the fits to data.

The detector response uncertainties associated with PID, photon detection both from the transition photon and from $\pi^{0}$/$\eta$ decay, and tracking of charged particles are approximately 2.5-2.9\% in every mass bin.

The uncertainty on $\epsilon_{38}$ from the fragmentation study is taken from the change in $\epsilon_{38}$ when modifying the weights given in Table~\ref{tab:ratios} by the uncertainty on these values individually.  We also account for the differences in statistics between the mass regions over which these uncertainties were determined and the individual mass bins. Since our fragmentation study procedure groups bins together before evaluating appropriate weights, the weights we identify tend to reflect the bins with higher numbers of events, and the uncertainty on the bins with fewer events needs to be increased.  We therefore increase the uncertainty in each $m_{X_{s}}$ bin by a factor of $\sqrt{N_{\rm region}}/\sqrt{N_{\rm bin}}$, where $N_{\rm region}$ ($N_{\rm bin}$) refers to the number of events in the region (bin).  This correction ensures that if an $m_{X_{s}}$ bin has few events compared to its corresponding region, then the uncertainty for this bin will be larger.    The total fragmentation uncertainty is found by summing in quadrature the changes for each of the ten subset amounts.  Where asymmetric uncertainties are reported in Table~\ref{tab:ratios}, we take the average change in $\epsilon_{38}$ when fluctuating the weights by the indicated amounts.  For the mass bin $1.0<m_{X_{s}}<1.1\gevcc$, it is unknown if the fragmentation in the data is modeled more effectively by the $K^{*}(892)$ MC sample or the inclusive MC sample.  We take the average of the two predictions to be the value for $\epsilon_{38}$, and the uncertainty is the difference divided by $\sqrt{12}$, consistent with the standard deviation of a uniform distribution.

The uncertainty on the missing fraction was covered in Sec.~\ref{sec:fragmentation} for the inclusive region.  The competing fragmentation models give an uncertainty on the missing fraction from 1.3 to 32.7\%, getting larger at higher mass.  For the $K^{*}(892)$ region, we take the uncertainty to be the difference between the default $K^{*}(892)$ MC prediction for the missing fraction, and the hypothesis of exclusively missing \KL final states, which would be a missing fraction of 25\% for this region.

We take each of these systematic uncertainties to be uncorrelated within an $m_{X_{s}}$ bin.  However, there are correlations in the errors between the mass bins.  The \BB\ counting, classifier requirements, non-\BB\ fitting for signal and cross-feed PDF shape, and detector response systematic uncertainties are taken to be completely correlated between all mass bins.  As we parameterize the peaking \BB\ Novosibirsk function in three different regions, we evaluate the uncertainties over the same regions, taking the uncertainties to be independent from one region to the other (indicated by the horizontal lines in Table~\ref{tab:systematic}).  Similarly, the fragmentation uncertainty and missing fraction uncertainty are evaluated using different samples and strategies in different mass regions; we take the uncertainty on these mass regions to be uncorrelated with one another, but completely correlated between the mass bins within a mass region. 


\begin{table*}[htp]
\begin{center}
\caption{List of systematic uncertainties described in the text.  These sub-component systematic uncertainties are assumed to be uncorrelated within a given mass bin and the total uncertainty reflects their addition in quadrature.  All uncertainties are given in \%.  Many of these uncertainties are taken to be completely correlated over $m_{X_{s}}$ regions, and we have indicated the correlated uncertainties with horizontal lines defining the regions.\label{tab:systematic}}
\begin{tabular}{ccccccccc}
\hline
\hline \\[-2.3ex]
Mass bin		& \BB			& Classifier		& Non-\BB	&\BB		& Detector	& Frag.		& Missing 	& Total\\
(\gevcc)	    & counting		& selection			& fitting	& fitting	& response		& 		& fraction	& \\
\hline
0.6--0.7			& 1.1			& 1.0				& 14.9		&21.3		& 2.5	& --		& 0.6			 &26.2\\
0.7--0.8			& 1.1			& 1.0				& 2.7   	&3.1		& 2.6	& --		& 0.9			 &5.1\\
0.8--0.9			& 1.1			& 1.0				& 1.7		&0.6		& 2.6	& --		& 1.3			 &3.8\\
0.9--1.0			& 1.1			& 1.0				& 1.7		&0.7		& 2.7	& --		& 0.0			 &3.6\\
1.0--1.1			& 1.1			& 1.0				& 5.1		&2.5		& 2.7	& 13.1		& 0.9			 &14.6\\
\cline{7-7}\cline{8-8}
1.1--1.2			& 1.1			& 0.7				& 5.7		&0.9		& 2.7	& 3.9		& 1.3			 &7.7\\
1.2--1.3			& 1.1			& 0.7				& 4.7		&0.4		& 2.7	& 3.0		& 1.3			 &6.4\\
1.3--1.4			& 1.1			& 0.7				& 4.6		&0.3		& 2.7	& 3.0		& 1.6			 &6.4\\
1.4--1.5			& 1.1			& 0.7				& 4.7		&0.6		& 2.7	& 5.7		& 1.8			 &8.2\\
\cline{7-7}
1.5--1.6			& 1.1			& 0.7				& 3.7		&1.5		& 2.7	& 6.1		& 3.1			 &8.5\\
1.6--1.7			& 1.1			& 0.7				& 4.3		&1.3		& 2.7	& 6.3		& 5.9			 &10.2\\
1.7--1.8			& 1.1			& 0.7				& 4.9		&1.5		& 2.7	& 7.9		& 6.9			 &12.1\\
1.8--1.9			& 1.1			& 0.7				& 3.4		&13.1		& 2.7	& 10.0		& 9.6			 &19.6\\
1.9--2.0			& 1.1			& 0.7				& 5.3		&4.2		& 2.7	& 13.4		& 11.1			 &18.9\\
\cline{5-5}\cline{7-7}
2.0--2.2			& 1.1			& 1.9				& 4.5		&6.6		& 2.9	& 11.0		& 13.9			 &19.8\\
2.2--2.4			& 1.1			& 1.9				& 4.9		&22.0		& 2.9	& 18.4		& 19.7			 &35.3\\
\cline{5-5}\cline{7-7}
2.4--2.6			& 1.1			& 2.8				& 4.7		&23.8		& 2.8	& 36.7		& 26.8			 &51.7\\
2.6--2.8			& 1.1			& 2.8				& 49.3		&154.1		& 2.8	& 45.7		& 32.7			 &171.3\\
\hline
\hline
\end{tabular}
\end{center}
\end{table*}


\section{Branching Fractions}

We measure the signal yield in 100 \mevcc wide bins of the $X_{s}$ mass over the range $0.6<m_{X_{s}}<2.0$\gevcc, and 200 \mevcc wide bins over the mass range $2.0<m_{X_{s}}<2.8$ \gevcc.  We report the measured signal yield in Table~\ref{tab:measuredSY}, where we have included the $\chi^{2}$ per degree of freedom (dof) from the fits.
\begin{table}[htp]
\begin{center}
\caption{Signal yields from fits to the on-peak data and corresponding $\chi^{2}$/dof from the fits (the uncertainties are statistical only).\label{tab:measuredSY}}
\begin{tabular}{crccc}
\hline
\hline
$m_{X_{s}}$     & \multicolumn{3}{c}{$N_{\mathrm{yield}}$}       & Data fit\\
(\gevcc)        & \multicolumn{3}{c}{(events)}    & $\chi^{2}$/dof\\
\hline
0.6--0.7         & 5.9   &$\pm$&12.2      & 0.8\\
0.7--0.8         & 114.7 &$\pm$&24.0    & 0.9\\
0.8--0.9         & 2627.4&$\pm$&50.2   & 1.0\\
0.9--1.0         & 2249.5&$\pm$&53.1   & 0.9\\
1.0--1.1         & 380.4 &$\pm$&36.1    & 0.9\\
1.1--1.2         & 393.7 &$\pm$&37.1    & 0.8\\
1.2--1.3         & 1330.5&$\pm$&47.1   & 0.6\\
1.3--1.4         & 1501.0&$\pm$&54.7   & 1.0\\
1.4--1.5         & 1479.6&$\pm$&58.3   & 1.0\\
1.5--1.6         & 1039.6&$\pm$&55.7   & 0.9\\
1.6--1.7         & 929.1 &$\pm$&56.7    & 0.9\\
1.7--1.8         & 736.5 &$\pm$&48.6    & 1.2\\
1.8--1.9         & 585.8 &$\pm$&50.8    & 1.0\\
1.9--2.0         & 272.0 &$\pm$&37.4    & 1.0\\
2.0--2.2         & 684.4 &$\pm$&68.2    & 1.1\\
2.2--2.4         & 277.5 &$\pm$&64.6    & 1.0\\
2.4--2.6         & 159.7 &$\pm$&54.4    & 0.8\\
2.6--2.8         & -34.4 &$\pm$&62.0    & 1.1\\
\hline
\hline
\end{tabular}
\end{center}
\end{table}

\begin{figure}[htp]
\begin{center}
\includegraphics[width=0.45\textwidth]{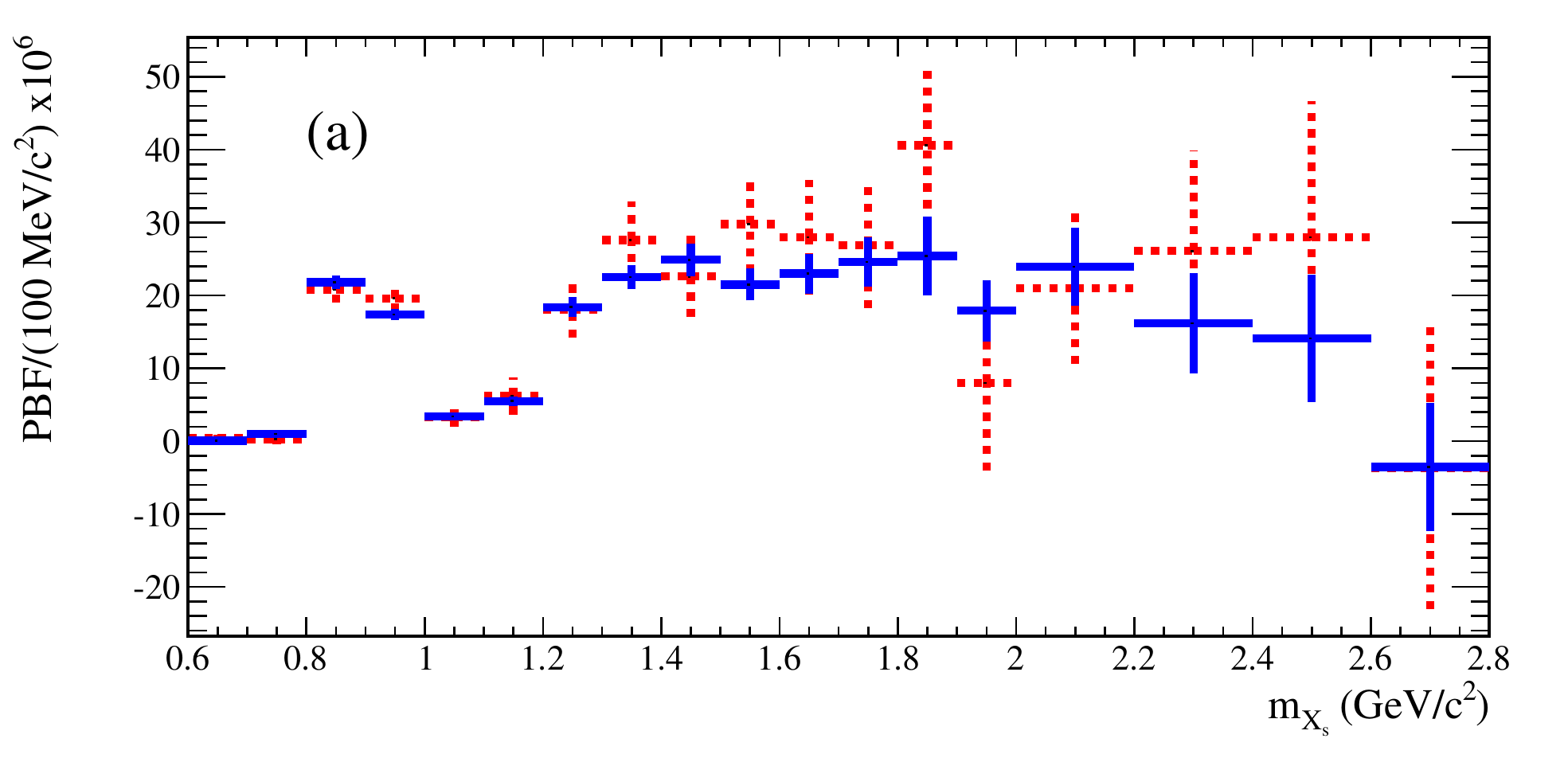}\\
\includegraphics[width=0.45\textwidth]{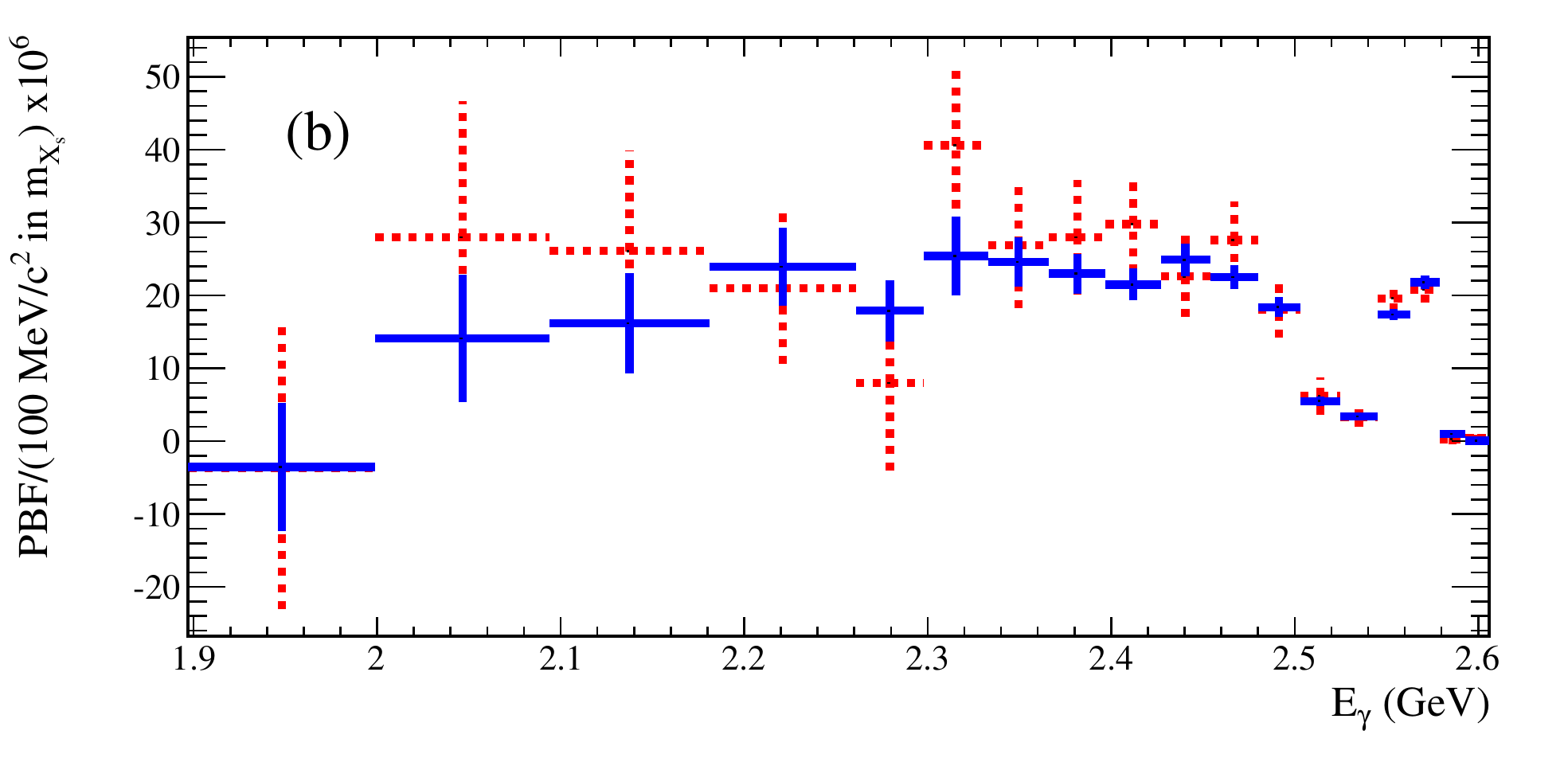}
\caption{The partial branching fractions binned in (a) $X_{s}$ mass and (b) the corresponding $E_{\gamma}$ bins, with the statistical and systematic uncertainties added in quadrature. The current results (solid lines) and former \babar\ results~\cite{babar_measurement} (dashed lines) are shown.\label{fig:mass_spectrum}}
\end{center}
\end{figure}

We use the efficiencies reported in Tables~\ref{tab:eps38} and~\ref{tab:incl} to derive the total number of \btosgs events, $N_{b\rightarrow s\gamma}$, based on the yields, $N_{\mathrm{yield}}$, reported in Table~\ref{tab:measuredSY}, according to:
\begin{equation}
N_{b\rightarrow s\gamma}=\frac{N_{\mathrm{yield}}}{\epsilon_{38}\epsilon_{\mathrm{incl}}}.
\end{equation}
\noindent The partial branching fraction (PBF) for each mass bin is reported in Table~\ref{tab:BFs}.  In this table, we also report the total branching fraction, with a minimum photon energy of $E_{\gamma}>1.9\gev$, reflecting the sum of the 18 bins:
\begin{equation}
 \mathcal{B}(\Bbar\rightarrow X_{s}\gamma)=(3.29\pm0.19\pm0.48)\times 10^{-4},
\end{equation}
where the first uncertainty is statistical and the second is systematic.
This result is consistent with the previous \babar\ sum of exclusives results of $\mathcal{B}(\Bbar\rightarrow X_{s}\gamma)=(3.27\pm0.18^{+0.55+0.04}_{-0.40-0.09})\times10^{-4}$~\cite{babar_measurement}, where the first uncertainty is statistical,
the second systematic, and the third from theory.  
The total statistical uncertainty on our result reflects the sum in quadrature of the statistical uncertainty of the 18 uncorrelated statistical uncertainties in the mass bins.  This method ensures reduced spectrum model dependence when quoting a branching fraction. An alternate method of measuring the transition rate based on larger mass bins yields similar results.  This alternative method is similar to the method used in the previous analysis~\cite{babar_measurement}, in which one measurement of the signal yield over the entire mass range was used to determine the total transition rate.  However, that method introduces additional model dependence due to the uncertainty in the spectrum shape and we instead decide to take the total transition rate as the sum of the transition rates in each of the $m_{X_{s}}$ bins.  The total systematic uncertainty reported in our study takes the correlations, indicated in Table~\ref{tab:systematic}, into account.  The correlation coefficients between the total uncertainties in each bin are included in the Appendix. The partial branching fractions per 100 \mevcc in $X_{s}$ mass are illustrated in Fig.~\ref{fig:mass_spectrum}, with the previous \babar\ sum-of-exclusive results also shown.

\begin{table}[htp]
\begin{center}
\caption{The partial branching fractions in each mass bin reflecting branching fractions per 100 or 200~\mevcc, and the total branching fraction for \btosgs with $E_{\gamma}>1.9\gev$.  The uncertainties quoted are statistical and systematic.\label{tab:BFs}}
\begin{tabular}{crcccc}
\hline
\hline
$m_{X_{s}}$     & \multicolumn{5}{c}{Branching fraction}\\
(\gevcc)        & \multicolumn{5}{c}{per 100 or 200~\mevcc}\\
                & \multicolumn{5}{c}{($\times10^{-6}$)}\\
\hline
0.6--0.7         & 0.1 &$\pm$&0.1&$\pm$&0.0\\
0.7--0.8         & 1.0 &$\pm$&0.2&$\pm$&0.1\\
0.8--0.9         & 21.8&$\pm$&0.4&$\pm$&0.8\\
0.9--1.0         & 17.4&$\pm$&0.4&$\pm$&0.6\\
1.0--1.1         & 3.4 &$\pm$&0.3&$\pm$&0.5\\
1.1--1.2         & 5.5 &$\pm$&0.5&$\pm$&0.4\\
1.2--1.3         & 18.4&$\pm$&0.7&$\pm$&1.2\\
1.3--1.4         & 22.5&$\pm$&0.8&$\pm$&1.5\\
1.4--1.5         & 24.9&$\pm$&1.0&$\pm$&2.0\\
1.5--1.6         & 21.5&$\pm$&1.2&$\pm$&1.8\\
1.6--1.7         & 23.0&$\pm$&1.4&$\pm$&2.3\\
1.7--1.8         & 24.6&$\pm$&1.6&$\pm$&3.0\\
1.8--1.9         & 25.4&$\pm$&2.2&$\pm$&5.0\\
1.9--2.0         & 17.9&$\pm$&2.5&$\pm$&3.4\\
2.0--2.2         & 24.0&$\pm$&2.4&$\pm$&4.7\\
2.2--2.4         & 16.2&$\pm$&3.8&$\pm$&5.7\\
2.4--2.6         & 14.1&$\pm$&4.8&$\pm$&7.3\\
2.6--2.8         & -3.5&$\pm$&6.4&$\pm$&6.1\\
\hline
0.6--2.8         & 329&$\pm$&19&$\pm$&48\\
\hline
\hline
\end{tabular}
\end{center}
\end{table}

\section{Fits to Spectrum Models and Moments}

Since we measure the hadronic mass spectrum in bins of 100 or 200 \mevcc, we are able to fit directly different models of this spectrum to obtain the best-fit values of different HQET parameters.  We choose to fit two such classes of models: the kinetic model using an exponential distribution function~\cite{BBU}, and the shape function model, also using an exponential distribution function~\cite{LNP}.  The choice of distribution function is not expected to have a large impact on the values determined for the underlying HQET parameters for each model, but the parameters themselves are not immediately comparable between models (for example, the models are evaluated at different energy scales).

In order to fit the measured spectrum to these models, we need to take special account of the $K^{*}(892)$ resonance, as the models assume quark-hadron duality in their spectra. Consequently, the models smooth over this resonance.  We fit a relativistic Breit-Wigner~\cite{RBW} (RBW) to the $K^{*}(892)$ MC sample at the generator level to extract the parameters of this curve. Fits to the transition point between the RBW curve of the $K^{*}(892)$ resonance and the remaining spectrum indicate a value close to $m_{X_{s}}$=1.17 \gevcc, which we take to be the location of this transition.   Furthermore, we require that the integral of the RBW used to parameterize the $K^{*}(892)$ region ($m_{X_{s}}<1.17 \gevcc$) be equivalent to the integral of this region in the spectrum models.  For the hadronic mass bin containing the transition from the $K^{*}(892)$ resonance to the nonresonant-spectrum models ($1.1<m_{X_{s}}<1.2\gevcc$), we assign the value of the integral of the RBW up to the transition point ($1.10<m_{X_{s}}<1.17 \gevcc$) plus the integral of the spectrum model from the transition point to the bin boundary ($1.17<m_{X_{s}}<1.20 \gevcc$).

We perform a fit to the different spectrum models by minimizing the quantity
\begin{equation}\label{eq:chi2}
\chi^{2}=\sum_{i,j}\frac{\left({\rm PBF}_{\mathrm{th}}-{\rm PBF}_{\mathrm{exp}}\right)_{i}C^{-1}_{ij}\left({\rm PBF}_{\mathrm{th}}-{\rm PBF}_{\mathrm{exp}}\right)_{j}}{\sigma_{i}\sigma_{j}},
\end{equation}
\noindent where ${\rm PBF}_{\mathrm{th}}$ and ${\rm PBF}_{\mathrm{exp}}$ are the PBF predicted by the spectrum model in the mass bin and the PBF we measured in the mass bin, respectively.  The matrix $C^{-1}_{ij}$ is the inverse of the matrix of correlation coefficients between the uncertainties on bins $i$ and $j$, reported in Appendix~\ref{sec:appen},
having taken the correlated systematic uncertainties and uncorrelated statistical uncertainties into account. The $\sigma_{i}$ and $\sigma_{j}$ are the total uncertainties (statistical and systematic added in quadrature) on the branching fractions determined for bins $i$ and $j$.

We find the best HQET parameter values based on the measured hadronic mass spectrum for two quantities for each model we fit.  For the kinetic model, we fix the chromomagnetic operator ($\mu_{G}^{2}$) to 0.35 $\gev^{2}$, 
and the expectation values of Darwin ($\rho^3_D= 0.2\gev^3$) and spin-orbit ($\rho^3_{LS}=-0.09\gev^3$) terms;
we allow $m_{b}$ and $\mu_{\pi}^{2}$ to take values between 4.45 and 4.75\gevcc and 0.2 and 0.7$\gev^{2}$, respectively.  We have points on the $m_{b}$-$\mu_{\pi}^{2}$ plane at which the spectrum has been evaluated exactly.  These points are spaced every 0.05\gevcc for $m_{b}$ and every 0.05$\gev^{2}$ for $\mu_{\pi}^{2}$.  We interpolate the spectrum mass bin predictions between these points using

\begin{widetext}
\begin{equation}\label{eq:interpolation}
F(m_{b},\mu_{\pi}^{2})=A+B\times(m_{b}-4.45)+C\times(\mu_{\pi}^{2}-0.2)+D\times(m_{b}-4.45)(\mu_{\pi}^{2}-0.2),
\end{equation}
\end{widetext}
\noindent where we solve this equation for [$A, B, C, D$].  The values 4.45 and 0.2 in Eq.(\ref{eq:interpolation}) are changed to the different values for which we have exact spectra provided.  This strategy ensures continuity in the value of the spectrum predictions for each hadronic mass bin across the $m_{b}$-$\mu_{\pi}^{2}$ plane.

The shape function models use two variables to parameterize the spectrum, $b$ and $\Lambda$~\cite{LNP}, that may be converted to values of $m_{b}$ and $\mu_{\pi}^{2}$, evaluated at a single energy scale of $1.5\gev$.  Similar to the kinetic model fits, we interpolate between points on the $b$-$\Lambda$ plane at which we have exact spectrum predictions (for $2.0\leq b\leq5.0$ in increments of 0.25, and $0.4\leq\Lambda\leq0.9\gev$ in increments of 0.05$\gev$).

\begin{figure}[htp]
\begin{center}
\includegraphics[width=0.45\textwidth]{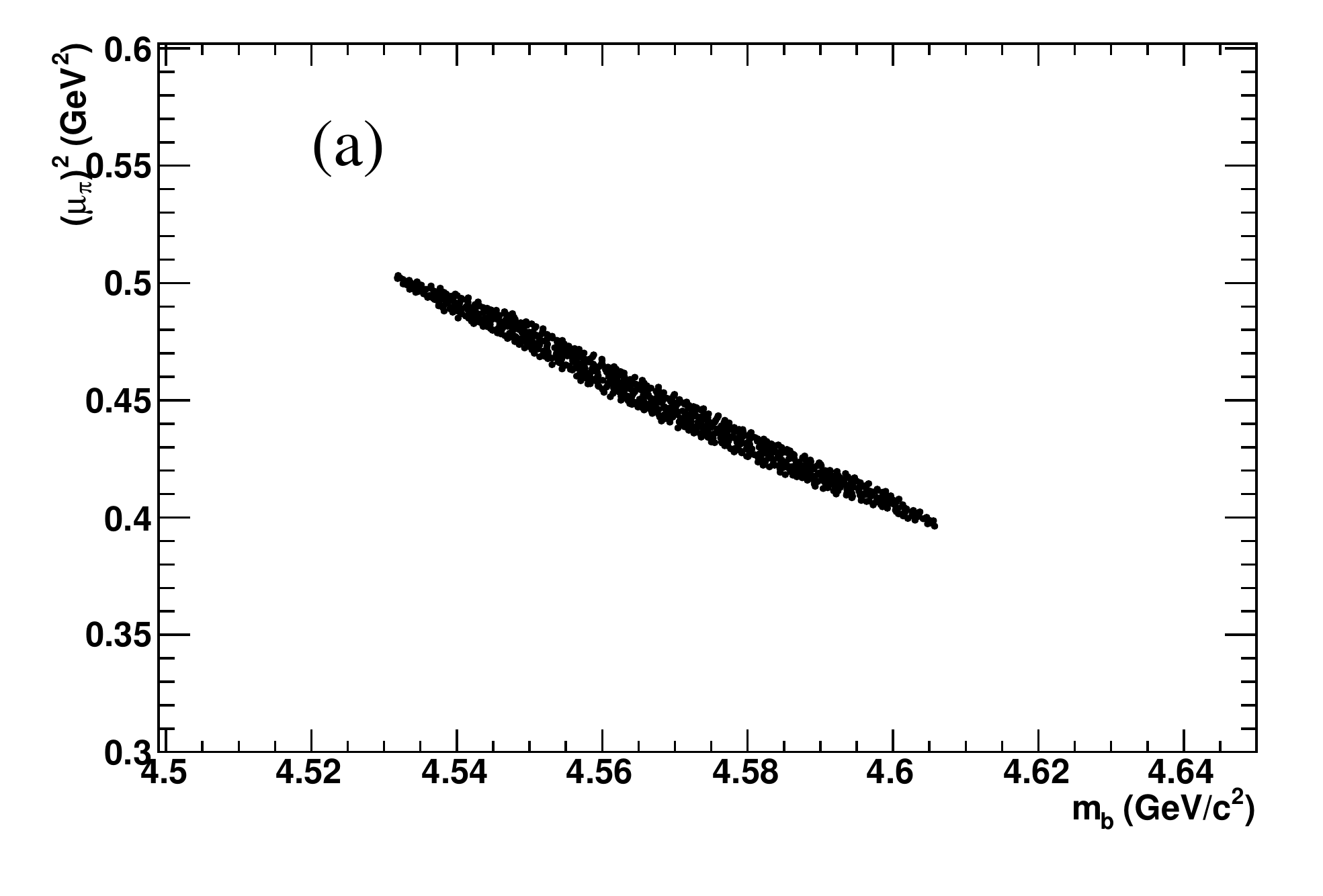}\\
\includegraphics[width=0.45\textwidth]{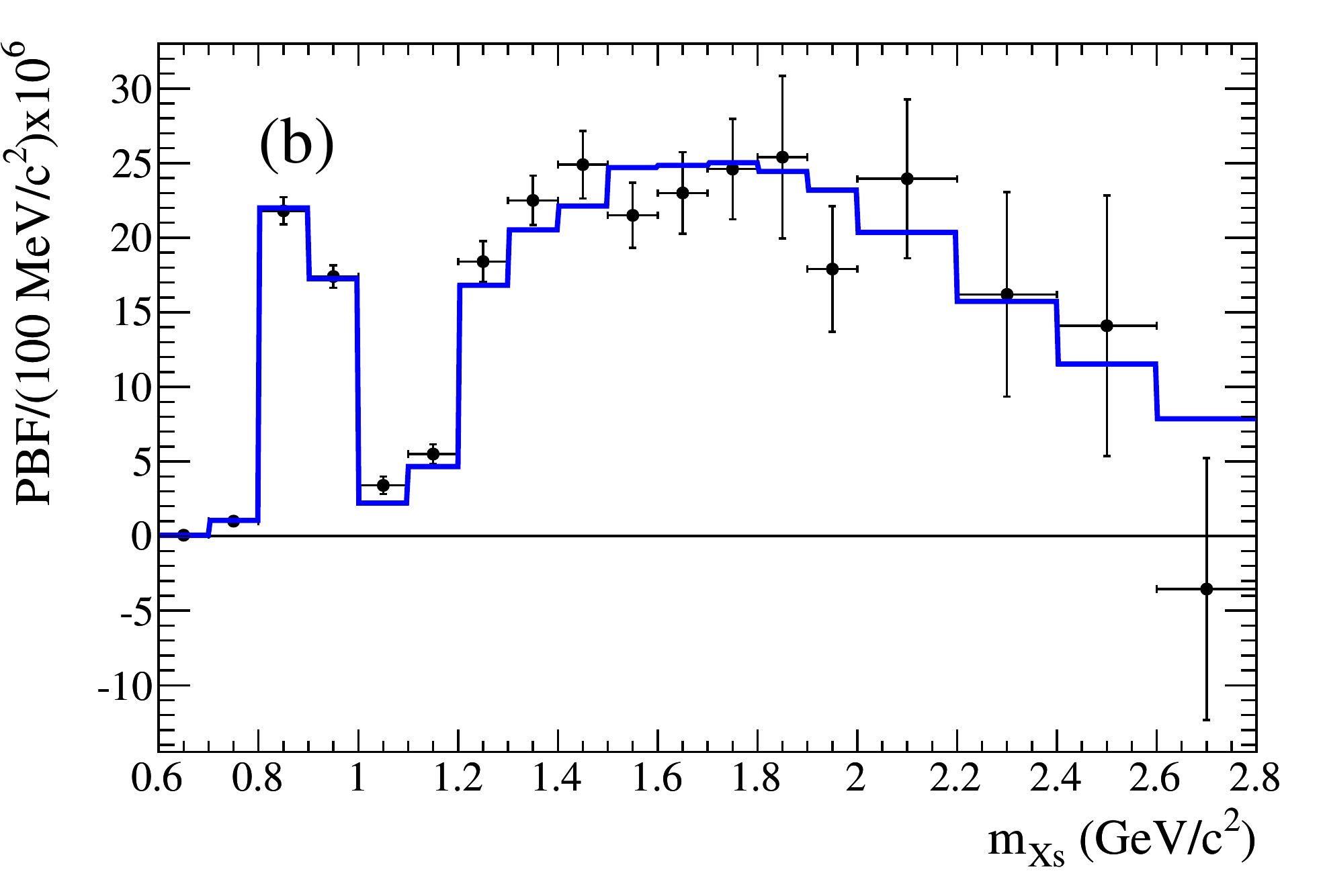}
\caption{The (a) One-$\sigma$ region for the kinetic model parameters based on the measured spectrum and the (b) best fit kinetic model compared to the measured PBFs.  The error bars in (b) include the statistical and systematic errors added in quadrature.\label{fig:BBU_result}}
\end{center}
\end{figure}

\begin{figure}[htp]
\begin{center}
\includegraphics[width=0.45\textwidth]{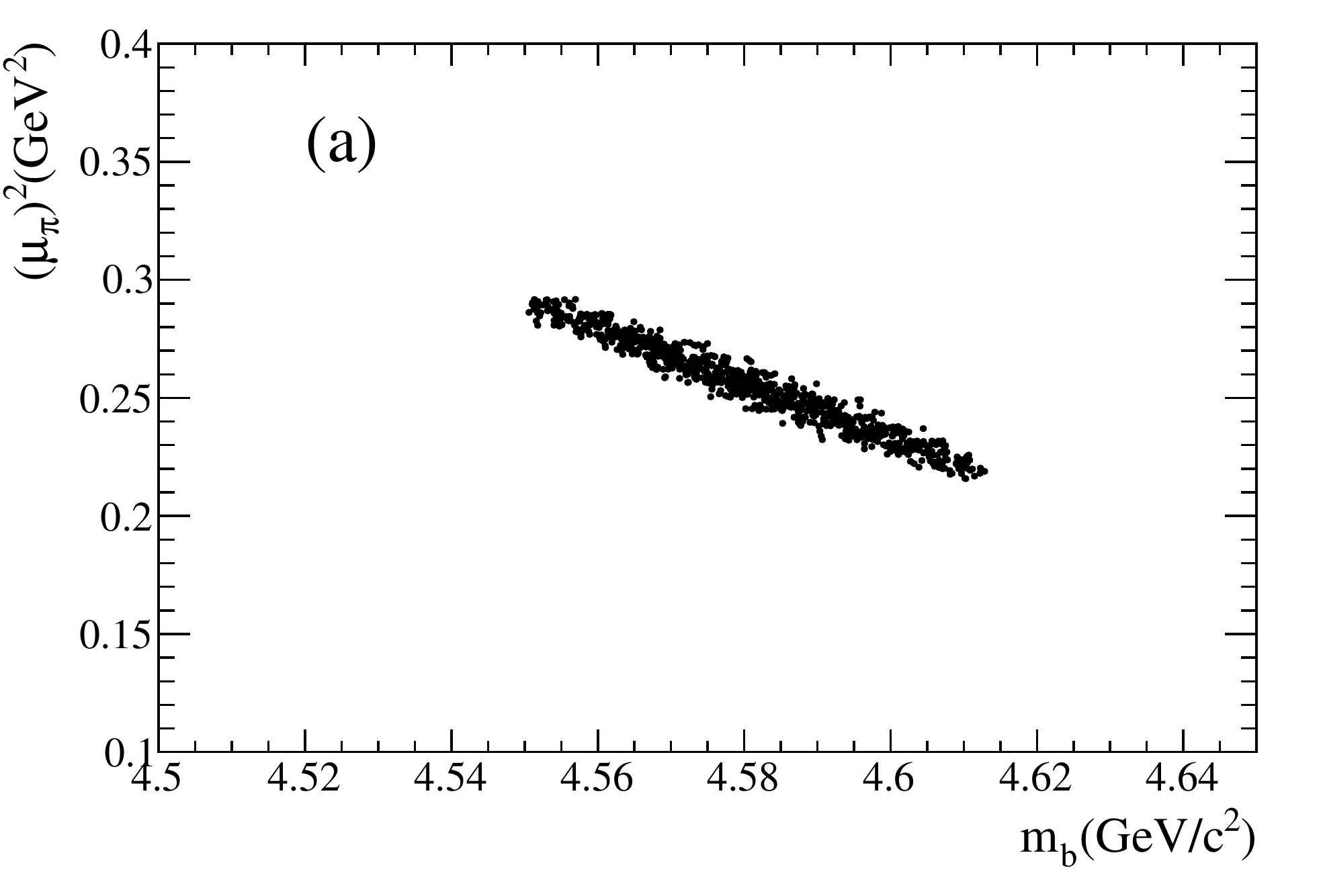}\\
\includegraphics[width=0.45\textwidth]{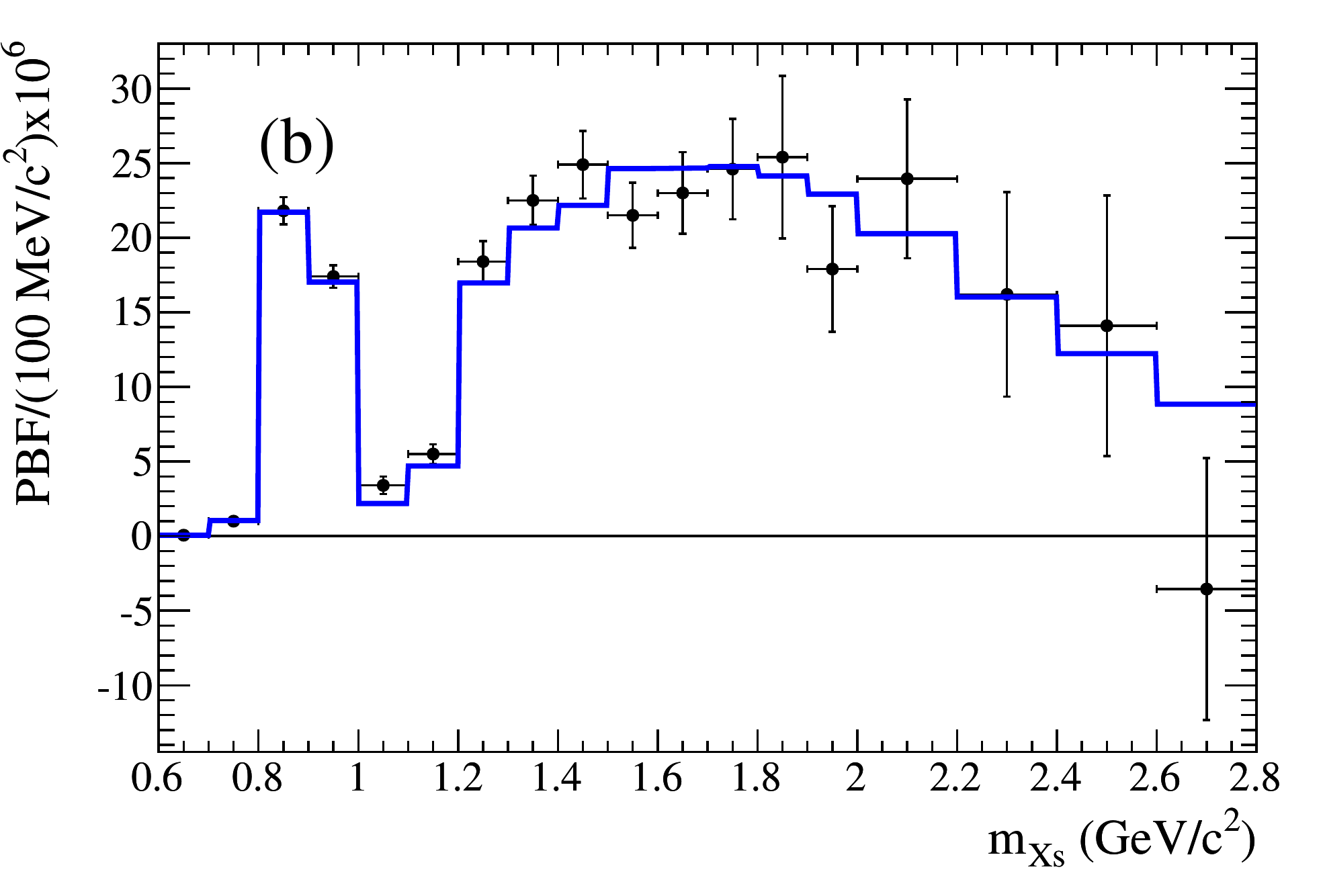}
\caption{The (a) One-$\sigma$ region for the shape function model parameters based on the measured spectrum and the (b) best fit shape function model compared to the measured PBFs.  The error bars in (b) include the statistical and systematic errors added in quadrature.\label{fig:SF_result}}
\end{center}
\end{figure}

The best fit values for the HQET parameters are reported in Table~\ref{tab:HQET_results}.  The uncertainty reflects the values at which the value of $\chi^{2}$ changes by one unit. The corresponding best fit spectrum model and $1\sigma$ error ellipses are shown in Fig.~\ref{fig:BBU_result} (kinetic model) and~\ref{fig:SF_result} (shape function model).


\begin{table}[htp]
\begin{center}
\caption{The best fit HQET parameter values based on the measured $m_{X_{s}}$ spectrum.\label{tab:HQET_results}}
\begin{tabular}{ccc}
\hline
\hline
                & Kinetic model~\cite{BBU}          & Shape function model~\cite{LNP} \\
\hline
\\[-2.3ex]
$m_{b}$         & $4.568^{+0.038}_{-0.036}\gevcc$   & $4.579^{+0.032}_{-0.029}\gevcc$ \\
\\[-2.3ex]
$\mu_{\pi}^{2}$ & $0.450\pm0.054\gev^{2}$ & $0.257^{+0.034}_{-0.039}\gev^{2}$\\
\hline
\hline
\end{tabular}
\end{center}
\end{table}

We use the PBFs measured in each mass bin to calculate the mean and variance of the photon energy spectrum.  These quantities are spectrum-model independent, and may be used to constrain the parameters in other models.  We evaluate the mean and variance for five different minimum photon energies and report the values in Table~\ref{tab:moments}.

\begin{table}[htp]
\begin{center}
\caption{The mean and variance of the photon energy spectrum, calculated for five photon energy cutoffs.  The errors are statistical and systematic.\label{tab:moments}}
\vspace{5mm}
\begin{tabular}{ccc}
\hline
\hline

      $E_{\gamma \mathrm{min}}$ & $\left<E\right>$ & $\left<E^{2}\right>-\left<E\right>^{2}$\\
	($\gev$)	& ($\gev$)	  &	($\gev^{2}$)\\

      \hline\\[-2.3ex]
      1.897            & 2.346$\pm$0.018$^{+0.027}_{-0.022}$ & 0.0211$\pm 0.0057^{+0.0055}_{-0.0069}$\\
\\[-2.1ex]
      1.999            & 2.338$\pm$0.010$^{+0.020}_{-0.017}$ & 0.0239$\pm 0.0018^{+0.0023}_{-0.0030}$\\
\\[-2.1ex]
      2.094            & 2.365$\pm$0.006$^{+0.016}_{-0.010}$ & 0.0176$\pm 0.0009^{+0.0009}_{-0.0016}$\\
\\[-2.1ex]
      2.181            & 2.391$\pm$0.003$^{+0.008}_{-0.007}$ & 0.0129$\pm 0.0003^{+0.0005}_{-0.0005}$\\
\\[-2.1ex]
      2.261            & 2.427$\pm$0.002$^{+0.006}_{-0.006}$ & 0.0082$\pm 0.0002^{+0.0002}_{-0.0002}$\\
\\[-2.1ex]
      \hline
\hline
\end{tabular}
\end{center}
\end{table}

We determine the pair-wise correlation between the uncertainties on the mean and variance calculated at different photon energy cutoffs.  We report these values in Table~\ref{tab:correlation}.  When determining the uncertainty on the means and variances, and evaluating the correlations between these uncertainties, we take into account the correlated systematic errors reported in Table~\ref{tab:systematic}.


\begin{table*}[htp]
\begin{center}
\caption{The correlation coefficients between $\left<E\right>$ and the variance for the different minimum photon energies based on the total uncertainties (statistical and systematic).\label{tab:correlation}}
\vspace{5mm}
\begin{tabular}{cccccccccccc}
\hline
	&	&\multicolumn{5}{c}{$\left<E\right>$}	 &\multicolumn{5}{c}{$\left<E^{2}\right>-\left<E\right>^{2}$} \\
	&	&\multicolumn{5}{c}{($\gev$)	}	 &\multicolumn{5}{c}{($\gev^{2}$)}\\
\hline
	&$E_{\gamma \mathrm{min}}$	& 1.897	& 1.999	& 2.094	& 2.181	& 2.261	& 1.897	& 1.999	& 2.094	& 2.181	& 2.261 \\
\hline
	&1.897	&1.00	&0.72	&0.46	&0.40	&0.20	&-0.90	&-0.66	&-0.36	&-0.27	&-0.13\\
	&1.999	&	&1.00	&0.71	&0.65	&0.35	&-0.39	&-0.29	&-0.52	&-0.40	&-0.18	\\
$\left<E\right>$	&2.094	&	&	&1.00	&0.84	&0.40	&-0.08	&-0.25	&-0.81	&-0.57	&-0.23\\
($\gev$)	&2.181	&	&	&	&1.00	&0.67	&-0.03	&-0.16	&-0.39	&-0.48	&-0.42\\
	&2.261	&	&	&	&	&1.00	&0.05	&0.04	&0.17	&0.31	&-0.68\\
\hline	
	&1.897	&	&	&	&	&	&1.00	&0.51	&0.13	&0.10	&0.00\\
	&1.999	&	&	&	&	&	&	&1.00	&0.29	&0.24	&-0.04\\
$\left<E^{2}\right>-\left<E\right>^{2}$	&2.094	&	&	&	&	&	&	&	&1.00	&0.69	&-0.10\\
($\gev^{2}$)	&2.181	&	&	&	&	&	&	&	&	&1.00	&-0.08\\
	&2.261	&	&	&	&	&	&	&	&	&	&1.00\\
\hline
\end{tabular}
\end{center}
\end{table*}

\section{Conclusion}

We have performed a measurement of the transition rate of \btosgs using the entire \babar\ \Y4S\ dataset.  We find that for $E_{\gamma}>1.9\gev$, the branching fraction is
\begin{equation}
\mathcal{B}(\Bbar\rightarrow X_{s}\gamma)=(3.29\pm0.19\pm0.48)\times 10^{-4},
\end{equation}
\noindent where the first uncertainty is statistical and the second is systematic.  The statistical uncertainty on this measurement is based on the sum in quadrature of the statistical uncertainties on each of the $X_{s}$ mass bin yields.  This method of combining statistical uncertainties ensures a reduced spectrum dependence and is different from
the method used in the previous \babar\ sum-of-exclusives approach where one large $m_{X_{s}}$ bin was used to determine the statistical uncertainty.  This measurement supersedes our previous measurement using the sum-of-exclusives approach~\cite{babar_measurement}.

We have also measured the mean and variance of the photon energy spectrum.  At the lowest photon energy cutoff ($E_{\gamma}>1.897\gev$), these values are
\begin{align}
\left<E\right> = 2.346\pm0.018^{+0.027}_{-0.022}\gev,\\
\left<E^{2}\right>-\left<E\right>^{2} = 0.0211\pm0.0057^{+0.0055}_{-0.0069}\gev^{2}.
\end{align}

Finally, we have also measured the best HQET parameters for two photon spectrum models.  For the shape function models~\cite{LNP} these are
\begin{align}
m_{b}=4.579^{+0.032}_{-0.029}\gevcc,\\
\mu_{\pi}^{2}=0.257^{+0.034}_{-0.039}\gev^{2}
\end{align}
\noindent (compared with the world averages of $m_{b}=4.588\pm 0.025\gevcc$ and $\mu_{\pi}^{2}=0.189^{+0.046}_{-0.057}\gev^{2}$~\cite{HFAG}), and for the kinetic models~\cite{BBU} these are
\begin{align}
m_{b}=4.568^{+0.038}_{-0.036}\gevcc,\\
\mu_{\pi}^{2}=0.450\pm0.054\gev^{2}
\end{align}
\noindent (compared with the world averages of $m_{b}=4.560\pm0.023\gevcc$ and $\mu_{\pi}^{2}=0.453\pm0.036\gev^{2}$~\cite{HFAG}). 
\section{acknowledgements}
We are grateful for the 
extraordinary contributions of our \pep2\ colleagues in
achieving the excellent luminosity and machine conditions
that have made this work possible.
The success of this project also relies critically on the 
expertise and dedication of the computing organizations that 
support \babar.
The collaborating institutions wish to thank 
SLAC for its support and the kind hospitality extended to them. 
This work is supported by the
US Department of Energy
and National Science Foundation, the
Natural Sciences and Engineering Research Council (Canada),
the Commissariat \`a l'Energie Atomique and
Institut National de Physique Nucl\'eaire et de Physique des Particules
(France), the
Bundesministerium f\"ur Bildung und Forschung and
Deutsche Forschungsgemeinschaft
(Germany), the
Istituto Nazionale di Fisica Nucleare (Italy),
the Foundation for Fundamental Research on Matter (The Netherlands),
the Research Council of Norway, the
Ministry of Education and Science of the Russian Federation, 
Ministerio de Ciencia e Innovaci\'on (Spain), and the
Science and Technology Facilities Council (United Kingdom).
Individuals have received support from 
the Marie-Curie IEF program (European Union) and the A. P. Sloan Foundation (USA).

\appendix
\section{Correlation Coefficients for Bin-Yield Uncertainties}\label{sec:appen}
In Table~\ref{tab:total_correlation_coefficient} we report the correlation coefficients between the total uncertainties reported in each mass bin.


\begin{center}
\begin{sidewaystable*}[htp]
\begin{small}
\caption{Correlation coefficients between the total uncertainties on the partial branching fractions measured in each mass bin.\label{tab:total_correlation_coefficient}}
\vspace{5mm}
\begin{tabular}{ccccccccccccccccccc}
$m_{X_{s}}$	& 0.6-0.7	& 0.7-0.8	&0.8-0.9	&0.9-1.0	&1.0-1.1	&1.1-1.2	&1.2-1.3	&1.3-1.4	 &1.4-1.5	 &1.5-1.6	&1.6-1.7	&1.7-1.8	&1.8-1.9	&1.9-2.0	&2.0-2.2	&2.2-2.4	 &2.4-2.6	 &2.6-2.8\\
(\gevcc) &&&&&&&&&&&&&&&&&&\\
\hline
0.6-0.7 &1.000  &0.025  &0.055  &0.056  &0.038  &0.045  &0.056  &0.054  &0.048  &0.046  &0.040  &0.039  &0.076  &0.036  &0.017  &0.009  &0.006  &-0.014\\
0.7-0.8 &0.025  &1.000  &0.182  &0.172  &0.083  &0.102  &0.141  &0.137  &0.117  &0.108  &0.094  &0.089  &0.125  &0.070  &0.047  &0.026  &0.018  &-0.026\\
0.8-0.9 &0.055  &0.182  &1.000  &0.697  &0.283  &0.378  &0.556  &0.544  &0.451  &0.387  &0.342  &0.314  &0.251  &0.207  &0.198  &0.107  &0.074  &-0.091\\
0.9-1.0 &0.056  &0.172  &0.697  &1.000  &0.268  &0.376  &0.552  &0.540  &0.449  &0.388  &0.342  &0.314  &0.267  &0.210  &0.196  &0.106  &0.073  &-0.089\\
1.0-1.1 &0.038  &0.083  &0.283  &0.268  &1.000  &0.192  &0.265  &0.257  &0.217  &0.183  &0.165  &0.157  &0.159  &0.114  &0.088  &0.048  &0.033  &-0.060\\
1.1-1.2 &0.045  &0.102  &0.378  &0.376  &0.192  &1.000  &0.552  &0.546  &0.551  &0.294  &0.293  &0.282  &0.203  &0.201  &0.202  &0.124  &0.096  &-0.111\\
1.2-1.3 &0.056  &0.141  &0.556  &0.552  &0.265  &0.552  &1.000  &0.762  &0.753  &0.423  &0.425  &0.409  &0.268  &0.288  &0.305  &0.189  &0.148  &-0.155\\
1.3-1.4 &0.054  &0.137  &0.544  &0.540  &0.257  &0.546  &0.762  &1.000  &0.751  &0.425  &0.436  &0.419  &0.276  &0.300  &0.325  &0.205  &0.164  &-0.156\\
1.4-1.5 &0.048  &0.117  &0.451  &0.449  &0.217  &0.551  &0.753  &0.751  &1.000  &0.361  &0.373  &0.360  &0.255  &0.263  &0.281  &0.178  &0.144  &-0.133\\
1.5-1.6 &0.046  &0.108  &0.387  &0.388  &0.183  &0.294  &0.423  &0.425  &0.361  &1.000  &0.703  &0.718  &0.611  &0.643  &0.317  &0.212  &0.180  &-0.119\\
1.6-1.7 &0.040  &0.094  &0.342  &0.342  &0.165  &0.293  &0.425  &0.436  &0.373  &0.703  &1.000  &0.751  &0.629  &0.674  &0.424  &0.293  &0.257  &-0.140\\
1.7-1.8 &0.039  &0.089  &0.314  &0.314  &0.157  &0.282  &0.409  &0.419  &0.360  &0.718  &0.751  &1.000  &0.645  &0.696  &0.421  &0.292  &0.257  &-0.139\\
1.8-1.9 &0.076  &0.125  &0.251  &0.267  &0.159  &0.203  &0.268  &0.276  &0.255  &0.611  &0.629  &0.645  &1.000  &0.642  &0.335  &0.238  &0.214  &-0.092\\
1.9-2.0 &0.036  &0.070  &0.207  &0.210  &0.114  &0.201  &0.288  &0.300  &0.263  &0.643  &0.674  &0.696  &0.642  &1.000  &0.364  &0.258  &0.230  &-0.109\\
2.0-2.2 &0.017  &0.047  &0.198  &0.196  &0.088  &0.202  &0.305  &0.325  &0.281  &0.317  &0.424  &0.421  &0.335  &0.364  &1.000  &0.701  &0.297  &-0.126\\
2.2-2.4 &0.009  &0.026  &0.107  &0.106  &0.048  &0.124  &0.189  &0.205  &0.178  &0.212  &0.293  &0.292  &0.238  &0.258  &0.701  &1.000  &0.216  &-0.085\\
2.4-2.6 &0.006  &0.018  &0.074  &0.073  &0.033  &0.096  &0.148  &0.164  &0.144  &0.180  &0.257  &0.257  &0.214  &0.230  &0.297  &0.216  &1.000  &0.057\\
2.6-2.8 &-0.014 &-0.026 &-0.091 &-0.089 &-0.060 &-0.111 &-0.155 &-0.156 &-0.133 &-0.119 &-0.140 &-0.139 &-0.092 &-0.109 &-0.126 &-0.085 &0.057  &1.000\\
\hline
\end{tabular}
\end{small}
\end{sidewaystable*}
\end{center}


\begin{thebibliography}{99}




\bibitem{2HDM}
G. Degrassi and P. Slavich, Phys. Rev. D \textbf{81}, 075001 (2010).

\bibitem{2HDM2}
F. Mahmoudi and O. St{\aa}l, Phys. Rev. D \textbf{81}, 035016 (2010).

\bibitem{2HDM3}
M. Ciuchini, G. Degrassi, P. Gambino, and G.F. Giudice, Nucl. Phys. B \textbf{534}, 3 (1998).


\bibitem{MFVMSSM}
M. Wick and W. Altmannshofer, AIP Conf.Proc. \textbf{1078}, 348 (2009).

\bibitem{UED}
A. Freitas and U. Haisch, Phys. Rev. D \textbf{77}, 093008 (2008).

\bibitem{NNLO}
M. Misiak and M. Steinhauser, Nucl. Phys. B \textbf{764}, 62 (2007); Nucl. Phys. B \textbf{840}, 271 (2010).

\bibitem{NNLO_second}
M. Misiak and M. Poradzinski, Phys. Rev. D \textbf{83}, 014024 (2011).

\bibitem{HFAG}
D. Asner \textit{et al.}, Heavy Flavor Averaging Group,``Averages of b-hadron, c-hadron, and tau-lepton Properties as of early 2012,'' arXiv:1207.1158 (2012).


\bibitem{Neubert}
M. Neubert, Phys. Rev. D \textbf{49}, 4623 (1994).

\bibitem{Wise}
A. Falk \textit{et al.}, Phys. Rev. D \textbf{49}, 4553 (1994).

\bibitem{Bauer}
C. Bauer and A. Manohar, Phys. Rev. D \textbf{70}, 034024 (2004).

\bibitem{Ligeti}
Z. Ligeti, I. Stewart and F. Tackmann, Phys. Rev. D \textbf{78}, 114014 (2008).

\bibitem{LNP}
B. Lange, M. Neubert and G. Paz, Phys. Rev. D \textbf{72}, 073006 (2005).

\bibitem{BBU}
D. Benson, I. Bigi and N. Uraltsev, Nucl. Phys. B \textbf{710}, 371 (2005).

\bibitem{babar_measurement}
B. Aubert \textit{et al.}, \babar\ Collaboration, Phys. Rev. D \textbf{72}, 052004 (2005).

\bibitem{BABARNIM}
B.\ Aubert {\em et al.}, \babar\ Collaboration, Nucl.~Instr.~and Methods A \textbf{479}, 1 (2002).

\bibitem{LST}
W. Menges \textit{et al.}, Nuclear Science Symposium Conference, 2005 IEEE, 3, 1470 (2005).


\bibitem{PDG}
K. Nakamura \textit{et al.} (Particle Data Group), J. Phys. G \textbf{37}, 075021 (2010).

\bibitem{KN}
A.L. Kagan and M. Neubert, Eur. Phys. J. C\textbf{7}, 5 (1999).

\bibitem{JETSET}
T. Sjostrand, Computer Physics Commun. \textbf{82}, 74 (1994).

\bibitem{geant}
S. Agostinelli \textit{et al.}, Nucl. Instrum. Methods A \textbf{506}, 250 (2003).

\bibitem{conjugates}
Charge conjugation is implied throughout this study.

\bibitem{ECOC}
T. G. Dietterich and G. Bakiri, Journal of Articifical Intelligence Research \textbf{2}, 263 (1995).

\bibitem{legendre}
B.\ Aubert {\em et al.}, \babar\ Collaboration, Phys. Rev. Lett. \textbf{89}, 281802 (2002).

\bibitem{RF}
L. Breiman, Machine Learning \textbf{45}, 5 (2001).

\bibitem{FW}
G. Fox and S. Wolfram, Nucl. Phys. B \textbf{149}, 413 (1979).

\bibitem{CB}
T. Skwarnicki, DESY internal report DESY-F31-86-02 (1986).

\bibitem{argus}
H. Albrecht \textit{et al.}, ARGUS Collaboration, Phys. Lett. B \textbf{241}, 278 (1990).

\bibitem{Nvs}
The Novosibirsk function is defined as
$f(m_{ES}) = \mathrm{exp}\left(-\frac{1}{2}\left(\ln^{2}[1+\Lambda\tau(m_{ES}-m)]/\tau^{2}+\tau^{2}\right)\right)$ where $\Lambda = \sinh(\tau\sqrt{\ln4})/(\sigma\tau\sqrt{\ln4})$, $m$ is the peak position, $\sigma$ is the width, and $\tau$ is the tail parameter.

\bibitem{QR}
C. Quigg and J. L. Rosner, Phys. Rev. D \textbf{17}, 239 (1978).

\bibitem{Fey}
R. D. Field and R. P. Feynman, Nucl. Phys. B \textbf{136}, 1 (1978).

\bibitem{RBW}
See Eq. (1) in B. Aubert \textit{et al.}, \babar\ Collaboration, Phys. Rev. D \textbf{78}, 071103 (2008).


\end{thebibliography}
\end{document}